\documentclass[prd,aps,preprint,tightenlines,superscriptaddress]{revtex4}
\usepackage{epsfig}
\usepackage{amsmath}
\def\m{\ln{\overline
N}}
\def\mm{\ln^2{\overline N}}
\def\mmm{\ln^3{\overline
N}}
\def\mmmm{\ln^4{\overline
N}}
\def\sl{\ln\left(\frac{Q^2}{\mu^2}\right)}
\def\ll{\ln^2\left(\frac{Q^2}{\mu^2}\right)}
\def\lll{\ln^3\left(\frac{Q^2}{\mu^2}\right)}
\def\llll{\ln^4\left(\frac{Q^2}{\mu^2}\right)}
\def\L{\tilde{\rm
L}}
\begin{document}

\preprint{DOE/ER/40762-361}

\title{Resummation of
Threshold Logarithms in Effective Field Theory
\\ for DIS, Drell-Yan and
Higgs Production }

\author{Ahmad Idilbi}
\email{idilbi@physics.umd.edu}
\affiliation{Department of Physics,
University of Maryland, College Park,
Maryland 20742, USA}
\author{Xiangdong Ji}
\email{xji@physics.umd.edu} \affiliation{Department of Physics,
University of Maryland, College Park, Maryland 20742, USA}
\affiliation{Department of Physics, Peking University, Beijing,
100871, P. R. China}
\author{Feng
Yuan}
\email{fyuan@quark.phy.bnl.gov} \affiliation{RIKEN/BNL
Research
Center, Building 510A, Brookhaven National Laboratory, Upton,
NY
11973}

\date{\today}
\vspace{0.5 in}
\begin{abstract}
We apply the effective field theoretic (EFT) approach to resum the
large perturbative logarithms arising when partonic hard
scattering cross sections are taken to the threshold limit. We
consider deep inelastic scattering, Drell-Yan lepton pair
production and the standard model Higgs production through
gluon-gluon fusion via a heavy-top quark loop. We demonstrate the
equivalence of the EFT approach with the more conventional,
factorization-based methods to all logarithmic accuracies and to
all orders in perturbation theory. Specific EFT results are shown
for the resummation up to next-to-next-to-next-to leading
logarithmic accuracy for the above-mentioned processes. We
emphasize the relative simplicity by which we derive most of the
results and more importantly their clear physical origin. We find
a new relation between the functions $f_{(q,g)}$ in the quark and
gluon form factors and the matching coefficients in Drell-Yan and
Higgs production, which may explain their universality believed to
hold to all orders in perturbation theory.
\end{abstract}

\maketitle

\section{Introduction}
Factorization theorems \cite{collins89} for inclusive hard
scattering processes are our main tool by which we quantitatively
analyze cross sections, with hadrons involved, when a generic hard
scale $Q^2$ is taken to infinity. Taking the Drell-Yan (DY)
lepton-pair production as an example, it is well known that the
cross section can be expressed as a convolution  of a hard
scattering coefficient function (or ``Wilson coefficient'')
calculated perturbatively, and a non-perturbative, universal,
parton distribution function (PDF) for each one of the incoming
hadrons. Corrections to the factorized cross section scale as
$1/(Q^2)^n$ where $n \geq 1$ up to some logarithmic ratios.
However, it is also well known that fixed order, pQCD calculation
of the Wilson coefficient yields singular distribution functions
of the form
\begin{eqnarray}
\alpha_s^k\left[\frac{\ln^{m-1}(1-z)}{(1-z)}\right]_+,\,\,\,\,
(m\leq 2k)
\end{eqnarray}
where $z=Q^2/{\hat s}$ and ${\hat s}$ is the total momentum
squared of the incoming partons. The ``plus'' distributions are
defined in the usual way. The appearance of such distributions is
a result of an emission of soft and/or collinear gluons into the
final state. When such distributions are Mellin transformed to the
conjugate space, logarithms of the form $\alpha_s^k\ln^m{\overline
N}$ ($m=2k,2k-1,...,0)$ show up where ${\overline N}\equiv
N\exp(\gamma_E)$ is the conjugate variable of $z$ and $\gamma_E$
is the Euler constant. In the limit $z\rightarrow 1$ or,
equivalently, large $\overline N$, fixed order perturbative calculation
cannot be reliably trusted and an all order resummation of the
large logarithms is needed. This is what is generically meant by
``threshold resummation''. This notion has evolved during the last
20 years into one of the most studied and highly developed
subjects within perurbative quantum chromodynamics (pQCD). Earlier
studies \cite{Ste87,{CatTre89}} supplied a sound and rigorous
(although complicated) treatment to perform such resummation. In
both of these works, resummation is performed after establishing
some sort of factorized cross section into well-defined quantities
(at the operator level) that capture the physics at the hard, jet
and soft scales. An integral transformation to the conjugate space
is then applied  in order to de-convolute the various terms in the
cross section. Then, in the conjugate space, energy evolution
equations are solved and the exponentials thus obtained contain
the resummed large logarithms. Thus the perturbative expansion is
put under control and the contributions obtained from the yet
uncalculated higher orders in $\alpha_s$ reduce the theoretical
uncertainty inherent in any fixed order calculation. Thereby
better phenomenological studies can be carried out and better
agreement with experimental data is usually obtained. More recent
studies have further developed and refined this topic
\cite{ster97,vogt1,ridolfi,kidonakis,vogel,RAVI}.

In this paper we adopt the effective field theory approach (EFT)
to resum the threshold large logarithms. This approach was first
applied to deep inelastic (non-singlet) structure function  in the
limit $x\rightarrow 1$ where $x$ is the partonic Bjorken variable
\cite{Man03}. Later on it was applied to the DY process in the
limit $z\rightarrow 1$ \cite{IdiJi05}. In both cases, resummation
was performed up to next-to-leading logarithms (NLL). The
implementation of the EFT methodology to resum threshold
logarithms is made more concrete due to the recently developed
``soft collinear effective theory'' (SCET) \cite{SCET,SCET1}.

The SCET describes interactions between soft and collinear
partons. It is the most appropriate framework to calculate
contributions from the soft-collinear limit of the full QCD
calculations (which is more commonly known as the ``soft limit'').
Therefore any perturbative calculations within SCET has to
reproduce the same results as the full QCD in that limit.  To
${\cal O}(\alpha_s)$, this has been verified explicitly for DIS
and DY \cite{Man03,IdiJi05}. This is also the case when one
considers distributions at small transverse momentum
\cite{Feng1,Gao}. Moreover, it was also shown that the one loop
diagrams (the form factor type of diagrams) calculated in SCET
have the same infrared (IR) pole structure as the full QCD
calculation \cite{Man03,Wei}. [The result in \cite{Man03} for the
collinear diagrams involve mixed poles of IR and ultraviolet (UV)
divergences. This would be treated by applying the ``zero-bin''
subtraction \cite{Man06}]. These observations have to be valid to
higher orders in the strong coupling. This allows us to extract
the relevant quantities needed to perform resummation from the
full QCD calculations as we shall see below.

The EFT resummation program described here is conceptually simple
and has been explained in detail in \cite{JiPRL}. The starting
point (and again considering  DY as an example) is the collinearly
factorized inclusive cross section in moment space
\cite{Catani:2003zt}
\begin{equation}
\label{sigmaN}
   \sigma_N =
\sigma_0\cdot G_N(Q)\cdot q(Q,N)\cdot q(Q,N),
\end{equation}
where $\sigma_0$ is the Born level cross section, $q(Q,N)$ is the
PDF of partons in hadrons and,
\begin{eqnarray}
\label{csfac}
   G_N(Q) &=&\vert
C(\alpha_s(Q^2))\vert^2e^{I_1(Q/\mu_I,\alpha_s(Q^2))}
     \times {\cal
M}_N(\alpha_s(Q^2))
e^{I_2(Q/\mu_I,\alpha_s(\mu_I^2))}e^{I_3(Q/\mu_I,\alpha_s(Q))}.
\end{eqnarray}
Explicit expressions for the various contributions in $G_N$ will
be given below; however, we want to comment on their physical
origin. $C(\alpha_s(Q^2))$ contains the non-logarithmic
contribution of the purely virtual diagrams and the first exponent
$I_1$ contains all the logarithms originating from the same type
of diagrams. Both quantities are obtained from the matching
procedure at the scale $Q$, and the running between $Q$ and the
intermediate scale $\mu_I$. This, in turn, is controlled by the
anomalous dimension of the EFT current to be denoted by
$\gamma_1$.

The intermediate scale $\mu_I$ shows up when real gluons are
emitted so one must consider the cross section with real gluon
emissions. The result will have both soft and collinear
divergences. When taking into account the IR poles from the
virtual diagrams (in the EFT approach this is done by taking into
account the contribution from the counterterms of the effective
operators), the total contribution will contain only collinear
divergences to be absorbed into a product of two PDFs. The
conclusion is that the matching procedure at the intermediate
scale is guaranteed to work to all orders in perturbation theory,
following the factorization theorem, as long as the EFT used
generates the full QCD results in the appropriate kinematical
limit and one gets the matching coefficient ${\cal
M}_N(\alpha_s(\mu_I))$ which by definition is finite in the
non-regulated theory. This quantity has to be free of any
logarithms. $I_2$ collects all the the logarithms that are due to
the evolution of the PDF between $\mu_I$ and the factorization
scale $\mu_F$. This is controlled by the anomalous dimension of
the PDF to be denoted by $\gamma_2$ . $I_3$ encodes all the
contributions due to the running of the coupling constant between
the matching scales ($Q$ and $\mu_I$) and the final factorization
scale $\mu_F$. All the large logarithms appear only in the
exponents and the term $\vert C(\alpha_s(Q^2))\vert^2{\cal
M}_N(\alpha_s(Q^2))$ is free of any large logarithms. In Eq.~(\ref
{csfac}) we have chosen $\mu_F=Q$ for simplicity.

The above formalism must be contrasted with the more conventional,
factorization based one. It will be shown that the EFT approach is
equivalent to the other approaches and to all logarithmic
accuracies. We will derive all the known ingredients needed to
perform threshold resummation to next-to-next-to-next-to leading
logarithmic accuracies (${\rm N}^3{\rm LL}$) for DIS non-singlet
structure function, DY process and the closely related Standard
Model (SM) Higgs production through gluon-gluon fusion into a top
quark loop. Moreover, the integrations in Eq.~(\ref{csfac}) are
very easy to perform and we have carried out the integrations up
to $g^{(3)}$ that resums the NNLL. This calculation is to be
compared with the ones explained in, e.g., Appendices A of
Refs.~\cite{Catani:2003zt,Vogt288}. In this paper we use
dimensional regularization in $d=4-2\varepsilon$ to regulate both
the UV and the IR divergences and we utilize the $\overline {\rm
MS}$ scheme throughout.

This paper is organized as follows. In Sec.II we derive the
anomalous dimension of the quark and gluon effective currents up
to ${\cal O}(\alpha_s^3)$ and write down the matching coefficients
at the scale $Q^2$ up to ${\cal O}(\alpha_s^2)$. In Sec.III we
obtain the matching coefficients at $\mu_I^2$ to ${\cal
O}(\alpha_s^2)$ and give our final expression for the resummed
coefficient function $G_N$. There we also comment on the
universality of the functions $f_{(q,g)}$ that enter the quark and
gluon anomalous dimensions of the effective operators. In Sec.IV
we compare the EFT approach with the conventional one and derive
our main result that establishes the full equivalence of the two
approaches. From that relation we obtain the recently calculated
$D^{(3)}_{(q,g)}$ for DY and Higgs production and ${\cal
B}^{(3)}_q$ for DIS. We carry out the integration in the resummed
coefficient function to illustrate the simplicity of the EFT
results and obtain the well-known functions
$g^{(i)}(\lambda)_{(q,g)}$ for $i=1,2,3$. Our conclusions are
presented in Sec.V. In the Appendix we write down explicit
expressions for soft and virtual limit in full QCD for all the
processes we consider up to ${\cal O}(\alpha_s^2)$ in $z$ space
and in the conjugate space (for large moments).

\section{Anomalous Dimension and Matching Coefficients for
Effective Currents}

The EFT approach for resummation starts from calculating the
contributions at scale $Q^2$. Technically this is done by matching
the full QCD theory currents to the EFT currents at the scale
$Q^2$ by considering the purely virtual diagrams in the full
theory. By doing this, we integrate out the hard modes of
virtualities of order $Q^2$. The matching of the currents can be
expressed as an operator expansion
\begin{equation}
   J_{\rm
QCD}=C(Q^2/\mu^2,\alpha_s(\mu^2)) J_{\rm eff}(\mu) + ...,
\end{equation}
where $C$ is the matching coefficient, $\mu$ is the factorization or
renormalization scale of the effective current and ellipses denote
higher-dimensional currents which will be ignored in this work. We
will consider the quark vector current $J^\mu = \bar \psi
\gamma^\mu\psi$ for DIS and DY cases and the gluon scalar current
$J=G^{\mu\nu}G_{\mu\nu}$ for Higgs production in hadron colliders.

The anomalous dimensions of the effective currents that control the
running (with $\mu$) are defined as
\begin{equation}
\label{evo1}
    \gamma_1 (\mu) = -\mu \frac{d\ln J_{\rm
eff}}{d\mu}.
\end{equation}
If the matrix elements of the currents in full QCD are independent
of the factorization scale, such as quark vector and axial vector
currents, the same anomalous dimensions are obtained from the
matching coefficients of the effective currents
\begin{equation}
    \gamma_1 (\mu) = \mu
\frac{d\ln C}{d\mu}.
\end{equation}
The anomalous dimension is a function of both $Q^2/\mu^2$ and
$\alpha_s(\mu^2)$. In fact, it can be shown that it is a linear
function of $\ln Q^2/\mu^2$ to all orders in perturbation theory
\cite{Man03};
\begin{equation}
      \gamma_1 = A(\alpha_s)\ln Q^2/\mu^2 +
B_1(\alpha_s), \label{adim}
\end{equation}
where $A$ and $B_1$ have expansions in
$a_s\equiv\alpha_s(\mu^2)/4\pi$: $A=\sum_ia_s^iA^{(i)}$ and
$B_1=\sum_ia_s^i B^{(i)}_1$, and $\alpha_s$ is the renormalized
coupling constant.

To obtain the anomalous dimensions and the matching coefficients, we
consider the simplest matrix element of the full QCD currents
between on-shell massless quark and gluon states. They are just the
on-shell form factors $F$. Since they are ``physical" observables,
there are no UV divergences, but there are IR ones. To all orders in
$\alpha_s$, we can write
\begin{equation}
\label{CS}
F= C(Q^2/\mu^2,\alpha_s(\mu^2))
S(Q^2/\mu^2,\alpha_s(\mu^2),1/\epsilon),
\end{equation}
where $S$ contains only infrared poles in dimensional
regularization (i.e., no finite terms). $S$ can be regarded as the
matrix element of the effective current $J_{\rm eff}$ after
renormalization has already been performed. In the effective
theory, Feynman diagrams for $S$ have vanishing contributions in
dimensional regularization because there are no scales in the
integrals. This can be regarded as the result of cancellation of
IR and UV poles. As such, the IR poles in $S$ may be treated as UV
poles for the purpose of calculating the anomalous dimension
\begin{equation}
    \gamma_1 (\mu) = -\mu \frac{d\ln
S}{d\mu}.
\end{equation}
Since $C$ does not contain any pole part, we can
also write
\begin{equation}
    \gamma_1 (\mu) = - \mu \left.\frac{d\ln
F}{d\mu}\right|_{\rm
    pole~ part}.
\end{equation}
Therefore, the perturbative results for $F$ up to any loop order can
be used to calculate the anomalous dimensions to the same order.

The best way to see the physical content of the form factor is to
consider a resummed form \cite{CollinsForm,Magnea90,Magnea2001}
\begin{equation}
\ln F(\alpha_s) = \frac{1}{2} \int^{Q^2/\mu^2}_0 \frac{d\xi}{\xi}
\left(K(\alpha_s(\mu),\epsilon)
   + G(1, \alpha_s(\xi\mu,\epsilon),
\epsilon) + \int^1_\xi
     \frac{d\lambda}{\lambda} A(\alpha_s(\lambda
\mu,\epsilon)\right),
\end{equation}
where $A$ is the anomalous dimension of the $K$ and $G$ functions,
\begin{equation}
   A(\alpha_s) =
\mu^2\frac{dG}{d\mu^2} = -
   \mu^2\frac{dK}{d\mu^2},
\end{equation}
and is in fact the same $A$ as in Eq.~(\ref{adim}). $K$ contains
only the IR poles, and therefore, the whole $K$-function can be
constructed from the perturbative expansion $A=\sum_i a_s^i
A^{(i)}$. The function $G$ contains only the hard contribution, and
has a perturbative expansion
\begin{eqnarray}
    G(1,\alpha_s,\epsilon) &=& \sum_i
    a_s^i
G^{(i)}(\epsilon).
\end{eqnarray}
Thus $\ln F$ can be expressed entirely
in terms of $G^{(i)}$ and
$A^{(i)}$.

The anomalous dimensions $A_q$ for the quark vector current (DIS and
DY) and $A_g$ for the gluon scalar current (Higgs production) have
been calculated up to ${\cal O}(\alpha_s^3)$ \cite{Vogt192},
\begin{eqnarray}
\label{a}
    A_{(q,g)}^{(1)} &=&
4C_{(q,g)}, \nonumber \\
    A_{(q,g)}^{(2)} &=& 8C_FC_{(q,g)}
\left[\left(\frac{67}{18}-\zeta_2\right)C_A
    -\frac{5}{9}N_F\right],
\nonumber \\
    A_{(q,g)}^{(3)} &=&
16C_{(q,g)}\left[C_A^2\left(\frac{245}{24} -
    \frac{67}{9}\zeta_2 +
\frac{11}{6}\zeta_3 +
    \frac{11}{5}\zeta^2_2\right) + C_F
N_F\left(-\frac{55}{24}  +
    2\zeta_3\right)\right. \nonumber \\
&&\left. + C_A N_F\left(-\frac{209}{108}+ \frac{10}{9}\zeta_2 -
\frac{7}{3} \zeta_3\right) +
N_F^2\left(-\frac{1}{27}\right)\right],
\end{eqnarray}
where $C_{(q,g)}=C_F$ for the quark and $C_A$ for the gluon. In this
sense  $A$ is universal.

The expansion coefficients for the $G$ function have been obtained
up to 3-loops from explicit calculations of the quark and gluon
form factors \cite{Vogt055}:
\begin{eqnarray}
G_{(q,g)}^{(1)} &=& 2(B_{2,(q,g)}^{(1)}- \delta_g\beta_0) +
f_{(q,g)}^{(1)} +
    \epsilon \tilde G_{(q,g)}^{(1)} + \epsilon^2 \tilde{\tilde
G}^{(1)}_{(q,g)}, \nonumber \\
    G_{(q,g)}^{(2)} &=& 2(B_{2,(q,g)}^{(2)}-
2\delta_g\beta_1) + f_{(q,g)}^{(2)} +
    + \beta_0 \tilde G_{(q,g)}^{(1)}+
\epsilon \tilde G_{q,g}^{(2)}, \nonumber \\
    G_{(q,g)}^{(3)} &=&
2(B_{2,(q,g)}^{(3)}- 3\delta_g\beta_2) + f_{(q,g)}^{(3)} +
    + \beta_1
\tilde G_{(q,g)}^{(1)}+ \beta_0 \left[\tilde
G_{(q,g)}^{(2)}-\beta_0\tilde {\tilde G}_{(q,g)}^{(1)}\right],
\label{gg}
\end{eqnarray}
where $\delta_g$ is zero for quark and 1 for gluon. The $B_2$'s
are the coefficients in front of the delta function $\delta (x-1)$
in the Altarelli-Parisi splitting function and have been
calculated to the third order \cite{Vogt192}:
\begin{eqnarray}
\label{b}
    B_{2,q}^{(1)} &=& 3C_F,
\nonumber \\
    B_{2,q}^{(2)} &=& 4C_FC_A\left(\frac{17}{24} +
\frac{11}{3}\zeta_2
    -3\zeta_3\right) - 4C_F N_F \left(\frac{1}{12} +
\frac{2}{3}\zeta_2\right)    + 4C_F^2\left(\frac{3}{8} - 3\zeta_2
+ 6\zeta_3\right),
    \nonumber \\
    B_{2,q}^{(3)} &=&
16C_AC_FN_F\left(\frac{5}{4} -
    \frac{167}{54}\zeta_2 +
\frac{1}{20}\zeta_2^2 +
    \frac{25}{18}\zeta_3\right)  \nonumber\\
    &&
+ 16C_AC_F^2\left(\frac{151}{64} + \zeta_2\zeta_3 -
\frac{205}{24}\zeta_2 - \frac{247}{60}\zeta_2^2 +
    \frac{211}{12}\zeta_3
+ \frac{15}{2}\zeta_5\right) \nonumber
    \\
    && - 16C_A^2C_F
\left(\frac{1657}{576} - \frac{281}{27}\zeta_2
    + \frac{1}{8}\zeta_2^2 +
\frac{97}{9}\zeta_3 -
    \frac{5}{2}\zeta_5\right) \nonumber \\
    && - 16
C_FN_F^2\left(\frac{17}{144} - \frac{5}{27}\zeta_2 +
\frac{1}{9}\zeta_3\right) \nonumber \\
    && - 16C_F^2 N_F
\left(\frac{23}{16} - \frac{5}{12}\zeta_2
      - \frac{29}{30}\zeta_2^2 +
\frac{17}{6}\zeta_3\right)
      \nonumber \\
    && +  16
C_F^3\left(\frac{29}{32} - 2\zeta_2\zeta_3
     + \frac{9}{8}\zeta_2 +
\frac{18}{5}\zeta_2^2 +
     \frac{17}{4}\zeta_3 -
15\zeta_5\right),
\end{eqnarray}
for quarks, and
\begin{eqnarray}
B_{2,g}^{(1)} &=& \frac{11}{3}C_A - \frac{2}{3}N_F,  \nonumber \\
B_{2,g}^{(2)} &=& 4C_AN_F \left(-\frac{2}{3}\right)
      + 4C_A^2
\left(\frac{8}{3} + 3\zeta_3\right)
       + 4C_F N_F
\left(-\frac{1}{2}\right),  \nonumber \\
    B_{2,g}^{(3)} &=& 16C_AC_FN_F
\left(-\frac{241}{288}\right)
     + 16C_AN_F^2\frac{29}{288} -
16C_A^2N_F\left(\frac{233}{288} + \frac{1}{6}\zeta_2
      +
\frac{1}{12}\zeta_2^2 + \frac{5}{3}\zeta_3\right)
      \nonumber \\
&& + 16C_A^3\left(\frac{79}{32} -\zeta_2\zeta_3
      + \frac{1}{6}\zeta_2 +
\frac{11}{24}\zeta_2^2
      + \frac{67}{6}\zeta_3 - 5\zeta_5\right)
\nonumber \\
      && + 16C_FN_F^2 \frac{11}{144} +
16C_F^2N_F\frac{1}{16},
\end{eqnarray}
for gluons.  The universal functions $f_{(q,g)}$ are given by
\begin{eqnarray}
\label{f}
   f_{(q,g)}^{(1)} &=& 0,  \nonumber
\\ f_{(q,g)}^{(2)} &=& C_{(q,g)}C_A \left[{808 \over 27} -{22 \over
3}\zeta_2
   -28\zeta_3\right]+
   C_{(q,g)} N_F\left[-{112 \over 27}+{8
\over 3}\zeta_2\right],  \nonumber \\
   f_{(q,g)}^{(3)} &=&
C_{(q,g)}C_A^2\left[\frac{136781}{729}-\frac{12650}{81}\zeta_2
-\frac{1361}{3}\zeta_3 + \frac{352}{5}\zeta^2_2 +
\frac{176}{3}\zeta_2\zeta_3 + 192\zeta_5\right]  \nonumber \\
    &&+C_{(q,g)}C_AN_F\left[-\frac{11842}{729} + \frac{2828}{81}\zeta_2
    +\frac{728}{27}\zeta_3 - \frac{96}{5}\zeta_2^2\right]
     +C_{(q,g)}
C_FN_F\left[-\frac{1771}{27} \right. \nonumber \\
     && \left.+ 4\zeta_2 +\frac{304}{9}\zeta_3 +
\frac{32}{5}\zeta_2^2\right]
      +
C_{(q,g)}N_F^2\left[-\frac{2080}{729} - \frac{40}{27}\zeta_2 +
\frac{112}{27}\zeta_3\right].
\end{eqnarray}
The tilted functions in Eq.~(\ref{gg}) are not given here since they
do not contribute to the anomalous dimension
as their contribution to the form factors are canceled (they can be found
in \cite{Vogt192}).

Finally, the anomalous dimension of the effective currents can be
expressed in terms of the $A$ and $G$ functions. If one writes
$\gamma_1 =\sum_ia_s^i \gamma_1^{(i)}$, then
\begin{equation}
\label{ano}
     \gamma_{1,(q,g)}^{(i)} =
A_{(q,g)}^{(i)} \ln Q^2/\mu^2
+B^{(i)}_{1,(q,g)}+2i\delta_g\beta_{i-1},
\end{equation}
where
\begin{equation}
B_{1,(q,g)}^{(i)}=-2B_{2,(q,g)}^{(i)}-f^{(i)}_{(q,g)}.
\end{equation}
and the QCD $\beta$-function is given by
\begin{eqnarray}
\beta(a_s)=-\frac{d\ln
\alpha_s}{d\ln
\mu^2}=\beta_0a_s+\beta_1a_s^2+....,
\end{eqnarray}
with $\beta_o=11C_A/3-2N_F/3$. The above expression for $\gamma_1$
might work to all orders in perturbation theory. In the gluon
case, the last term is present when the anomalous dimension is
defined in terms of the matching coefficient $C_g$ and is absent
when it is defined in term of the effective current.

The anomalous dimensions could also be calculated from the matching
coefficient $C(Q^2/\mu^2,\alpha_s(\mu^2))$ extracted from known
results of the form factors. First, we take the logarithm of
Eq.~(\ref {CS}),
\begin{equation}
\ln F=\ln C(Q^2/\mu^2)+\ln S(Q^2/\mu^2,1/\epsilon).
\end{equation}
Then we separate out the poles from the form factor logarithms,
which belong to the $S(Q^2/\mu^2,1/\epsilon)$. The finite part
left over is just the logarithm of the matching coefficient $\ln
C$ to any desired order. So, eventually, we will get the following
result for the anomalous dimension valid to arbitrary order in
$\alpha_s$,
\begin{equation}
\gamma_1=\frac{d}{d\ln\mu}\left\{\ln F|_{\rm finite~part}\right\},
\end{equation}
where the form factor in the above equation has been renormalized
(including coupling constant renormalization). Using the above
equation, we have calculated the anomalous dimension for the quark
and gluon currents in the effective theory up to three-loop order,
and they are exactly the same as Eq.~(\ref {ano}). We should point
out here that the anomalous dimension of the quark current is the
same for the scattering case (DIS) and for the annihilation case
(DY).

To calculate the matching coefficients,
$C(Q^2/\mu^2,\alpha_s(\mu^2))=\sum_ia_s^i(\mu^2)C^{(i)}(Q^2/\mu^2)$
for DIS, DY and Higgs production, we need the expressions for the
quark form factor (space-like case and time-like case)
\cite{Matsuura} , and for SM Higgs production
\cite{Sall,Catani01,Har01} up to the same order. It should be
noted that for DY and Higgs cases, $C^{(i)}$ contains imaginary
parts that need be taken into account. For our purposes, it is
enough to keep the imaginary part for $C^{(1)}$ only. Normalizing
$C^{(0)}$ to $1$ we find for DIS,
\begin{eqnarray}
C^{(1)}_{\rm
DIS}(Q^2/\mu^2)&=&C_F\left[-\ll+3\sl-8+\zeta_2\right],\nonumber
\\
C^{(2)}_{\rm
DIS}(Q^2/\mu^2)&=&
C_F^2\left[\frac{1}{2}\left(\ll-3\sl+8-\zeta_2\right)^2\right.\nonumber\\
&&\left.+
\left(\frac{3}{2}-12\zeta_2+24
\zeta_3\right)\sl-\frac{1}{8}+29\zeta_2-30\zeta_3-\frac{44}{5}\zeta_2^2\right]\nonumber\\
&&+C_FN_F\left[-\frac{2}{9}\lll+\frac{19}{9}\ll-\left(\frac{209}{27}+\frac{4}{3}\zeta_2\right)\sl\right.\nonumber\\
&&\left.+\frac{4085}{324}+\frac{23}{9}\zeta_2+\frac{2}{9}\zeta_3\right]\nonumber\\
&&+C_FC_A\left[\frac{11}{9}\lll+\left(2\zeta_2-\frac{233}{18}\right)\ll\right.\\
&&\left.+\left(\frac{2545}{54}+\frac{22}{3}\zeta_2-26\zeta_3\right)\sl-\frac{51157}{648}-\frac{337}{18}\zeta_2
+\frac{313}{9}\zeta_3+\frac{44}{5}\zeta_2^2\right].\nonumber
\end{eqnarray}
The logarithms in the above result have been presented in
\cite{Feng1}. For DY we can simply get the $C^{(i)}_{q}$ by
replacing each $\sl$ in $C^{(i)}_{\rm DIS}$ with $\sl-i\pi$. This is
just a result of the fact that the time-like quark form factor can
be obtained from the space-like one by analytic continuation. For
the Higgs production we set $M_H^2=Q^2$ and we get
\begin{eqnarray}
C^{(1)}_g(Q^2/\mu^2)&=&C_A\left[-\ll+7\zeta_2+2i\pi^2\sl
\right],\nonumber\\
{\rm Re}[C^{(2)}_g(Q^2/\mu^2)]&=&C_A^2\left[{1
\over
2}\llll+{11\over
9}\lll-\left({67 \over
9}-17\zeta_2\right)\ll\right.\nonumber\\
&&\left.+\left({80\over 27}-{88\over
3}\zeta_2-2\zeta_3\right)\sl+{5105 \over 162}+{335\over
6}\zeta_2-{143\over 9}\zeta_3+{125\over
10}\zeta_2^2\right]\nonumber\\
&&+C_AN_F\left[-{2\over9}\lll+{10\over
9}\ll+\left({52\over
27}+{16\over
3}\zeta_2\right)\sl\right.\nonumber\\
&&\left.-{916\over 81}-{25\over
3}\zeta_2-{46\over
9}\zeta_3\right]\nonumber\\
&&+C_FN_F\left[2\ll -{67\over
6}+8\zeta_3\right].
\end{eqnarray}
The logarithms in $C^{(i)}$ will be needed later on to show that
the matching coefficients at $\mu_I$ are free of any logarithms,
and we have not included the imaginary part of $C^{(2)}_g$ since
it does not contribute to the accuracy in which we are interested.

Using Eq.~(\ref{evo1}) we can write down the solution of the
renormalization group equations for DY and Higgs,
\begin{equation}
\label {ss} C_{(q,g)}(Q^2/\mu_I^2,\alpha_s(\mu_I^2)) =
C_{(q,g)}(1,\alpha_s(Q^2))~{\rm
exp}{\left[\frac{I_1(Q,\mu_I)}{2}\right]},
\end{equation}
where
\begin{eqnarray}
   I_1 &=&
-\int_{\mu_I}^Q\tilde \gamma_{1,(q,g)}
\frac{d\mu}{\mu} \nonumber \\
\tilde{\gamma}_{1,(q,g)}&=&\gamma_{1,(q,g)}-2i\delta_g\beta_{i-1}.
\end{eqnarray}
The $C(1,\alpha_s(Q^2))\equiv C(\alpha_s(Q^2))$ is just the
non-logarithmic part of $C(Q^2/\mu^2,\alpha_s(\mu^2))$. For the
Higgs case the last relation is a result of the $\mu$ dependence
of $C_\phi(\mu)$ which enters into the effective lagrangian that
one obtains after integrating out the top quark (see, e.g.,
\cite{Sall}). This $\mu$-dependence is governed by anomalous
dimension which we denote by $\gamma_T$ following the notation of
\cite{Feng1}. There it was shown that
\begin{equation}
\gamma_T=a_s[-2\beta_0]+a_s^2[-4\beta_1]\,\,,
\end{equation}
so the conclusion is that the only effect of this anomalous
dimension, when combined with anomalous dimension of the matching
coefficient at the scale $Q^2$ is to cancel the $\beta_i$ terms in
$\gamma_1$ for the Higgs case. For DIS we replace $C_q$ with $C_{\rm
DIS}$ which runs with the same $\gamma_{1,q}$.

In Eq.~(\ref {ss}) we encounter the first of three exponentials.
The other two will be obtained below. Since $\mu_I^2$ will later
be identified with $Q^2/{\overline N}^p$ where $p=1$ for DIS and
$p=2$ for DY and Higgs cases, it is clear that the exponential
includes large logarithms of the form mentioned in the
introduction. We again stress the fact that $C(\alpha_s(Q^2))$
(for all three processes) and $\gamma_{1,(q,g)}$ are completely
determined to a given ${\cal O}(\alpha_s^k)$ by the knowledge of
the form-factor calculation up to the same order.

\section{Matching Coefficients at $\mu_I$ and the Resummed
Coefficient Functions}

In this section we show how to extract the matching coefficients at
the intermediate scale to ${\cal O}(\alpha_s^2)$ for DIS, DY and
Higgs production from the known calculations of full QCD, and to
obtain resummed expressions for the coefficient functions. Since we
are interested in the threshold region, we need to consider only the
partonic channels that give rise to the singular contributions in
the limit $z\rightarrow 1$, i.e, $\delta(1-z)$ and the ``plus''
distributions, ${\cal D}_i(z)$, where
\begin{equation}
{\cal D}_i(z)\equiv\left[
\frac{\ln^i(1-z)}{1-z}\right]_+\,.
\end{equation}
For DIS, DY and the Higgs processes, these channels are:
$q+\gamma^*\rightarrow q$, $q+{\bar q}\rightarrow \gamma^*$ and
$g+g\rightarrow H$, respectively.

To the accuracy we are interested in, the ${\cal O}(\alpha_s^2)$
cross section from soft contributions are needed. The full QCD
calculations for cross sections can be found in
Refs.~\cite{Matsuura} for DY, in Refs.~\cite{Catani01,Har01} for
the Higgs production and in Refs.~\cite{Zil1,Zil2} for DIS. The
result in the soft limit can be written as
\begin{equation}
G^{(\rm {s+v})}(z)\equiv \sum_ia_s^i(\mu^2)G^{(i),(\rm {s+v})}(z)
\end{equation}
which contains both soft and virtual contributions, and where
$G(z)$ is the inverse Mellin transform of $G_N$ in
Eq.~(\ref{sigmaN}). Explicit expressions for $G^{(i),(\rm
{s+v})}(z)$ with $i=1,2$ can be found in Ref.~\cite{Zil3} for DIS,
in Ref.~\cite{Hamberg} for DY and in Ref.~\cite{Smith1,Smith2} for
the Higgs production.
 [$G^{(0)}(z)=\delta(1-z)$]. Using the
following well-known Mellin transforms of ${\cal D}_i(z)$ in the
large $N$ limit
\begin{eqnarray}
{\cal D}_0(\overline N)&=&-\ln {\overline
N}, \nonumber\\
{\cal D}_1(\overline N)&=& {1\over 2}\ln^2{\overline
N}+{1\over
2}\zeta_2,
\nonumber
\\
{\cal D}_2(\overline N)&=&-{1\over
3}\ln^3{\overline
N}-\zeta_2\ln{\overline N}-{2\over 3}\zeta_3,\nonumber
\\
{\cal D}_3(\overline N)&=&{1\over 4}\ln^4{\overline
N}+{3\over
2}\zeta_2\ln^2{\overline N}+2\zeta_3\ln{\overline
N}+{27\over
20}\zeta_2^2,
\end{eqnarray}
we get the $G^{(i),{\rm (s+v)}}(\overline N)$, $i=0,1,2$. Explicit
expressions are given in the Appendix. As we have already mentioned,
the SCET is supposed to reproduce the same results.

To get the matching coefficient at the intermediate scale $\mu_I$,
${\cal M}_N=\sum_ia_s^i{\cal M}^{(i)}_N$, we need to factorize the
virtual contribution from the following relation,
\begin{equation}
\label {master1}
G_N^{({\rm
s+v})}\left(\frac{Q^2}{\mu^2},{\overline
N},\alpha_s(\mu^2)\right)=\Big|
C\left(\frac{Q^2}{\mu^2},\alpha_s(\mu^2)\right)\Big|^2 \times
{\cal M}_N\left(\frac{Q^2}{\mu^2},{\overline
N},\alpha_s(\mu^2)\right).
\end{equation}
The content of this formula is simple: The finite part of the
partonic cross section $G_N^{({\rm s+v})}$ comes from both the
purely virtual, form-factor type of Feynman diagrams, which are
included in $\vert C\vert^2 $, and from diagrams with at least one
real gluon emitted into the final state, which are included in
${\cal M}_N$. However, there is a different way to look at it. The
right-hand side is just the result of a two-step matching of the
product of two full QCD currents where at each step we collect the
relevant contribution to the cross section. The first step
accounts for $\vert C\vert^2 $ and the second one gives rise to
${\cal M}_N$. It should be noted that multiple matching procedure,
as the one performed here, results in a multiplicative matching
coefficients. We also mention that the above equation could
formally be proved, inductively in $\alpha_s$, by considering the
cross section within the effective theory itself and relating it
to the full QCD calculation in the soft limit.

Expanding the above equation to the third order, one gets
\begin{eqnarray}
\label{s+v}
G^{(1),{\rm s+v}}_N&=&2{\rm
Re}[C^{(1)}]+{\cal M}^{(1)}_N, \nonumber\\
G^{(2),{\rm
s+v}}_N&=&|C^{(1)}|^2+2{\rm Re}[C^{(2)}]+2{\rm
Re}
[C^{(1)}]{\cal
M}_N^{(1)}+{\cal M}_N^{(2)},\\
G^{(3),{\rm s+v}}_N&=&2{\rm Re}[C^{(1)}C^{(2)\ast}]+2{\rm
Re}[C^{(3)}]+\vert C^{(1)}\vert^2{\cal M}_N^{(1)}\nonumber \\
&& + 2{\rm
Re}[C^{(1)}]{\cal M}_N^{(2)}+2{\rm Re}[C^{(2)}]{\cal
M}_N^{(1)}+{\cal M}_N^{(3)}.\nonumber
\end{eqnarray}
The above factorization is consistent with that considered in
\cite{MOCH}. We get the following result for DIS,
 \begin{eqnarray}
 {\cal M}_{{N,\rm
 DIS}}^{(1)}&=&C_F\left[2{\rm L}^2+3{\rm
 L}+7-4\zeta_2\right],\nonumber\\
 {\cal M}_{N,{\rm DIS}}^{(2)}&=&C_F^2\left[2{\rm L}^4+6{\rm
 L}^3+\left(\frac{37}{2}-8\zeta_2\right){\rm L}^2+\left({45\over
 2}-24\zeta_2+24\zeta_3\right){\rm L}\right]
 \nonumber\\
 &&+C_FC_A\left[\frac{22}{9}{\rm
 L}^3+\left(\frac{367}{18}-4\zeta_2\right){\rm
 L}^2-\left(-\frac{3155}{54}+\frac{22}{3}\zeta_2+40\zeta_3\right){\rm
 L}\right]\nonumber\\
 &&-C_FN_F\left[ \frac{4}{9}{\rm L}^3+\frac{29}{9}{\rm
 L}^2-\left(\frac{4}{3}\zeta_2-\frac{247}{27}\right){\rm
 L}\right]\nonumber\\
 &&+C_F^2\left[\frac{205}{8}-\frac{97}{2}\zeta_2-6\zeta_3+\frac{122}{5}\zeta_2^2\right]
 +C_FC_A\left[\frac{53129}{648}-\frac{155}{6}\zeta_2-18\zeta_3-\frac{37}{5}\zeta_2^2\right]\nonumber\\
 &&+C_F
 N_F\left[-\frac{4057}{324}+\frac{13}{3}\zeta_2\right],
 \end{eqnarray}
where ${\rm L}=\ln\frac{\mu^2{\overline N}}{Q^2}$. The above result has also been obtained in \cite {neu1} where
 an explicit two-loop calculation of a suitably defined jet function was performed \footnote {We thank the authors of Ref.~[39]
 for pointing out some misprints in Eq.~(34) that appeared in an earlier version of this paper.}. For DY, we get
\begin{eqnarray}
{\cal
M}_{N,q}^{(1)}&=&C_F\left[2{\L}^2+2\zeta_2\right],\nonumber\\
{\cal
M}_{N,q}^{(2)}&=&C_F^2\left({1\over
2}\right)\left[2{\L}^2+2\zeta_2\right]^2+C_AC_F\left[\frac{22}{9}{\L}^3+\left(\frac{134}{9}-4\zeta_2\right)
{\L}^2+\left(\frac{808}{27}-28\zeta_3\right)
{\L}\right]\nonumber\\
&&-C_FN_F\left[
\frac{4}{9}{\L}^3+\frac{20}{9}
{\L}^2+\frac{112}{27}{\L}\right]\nonumber\\
&&+C_F
C_A\left[\frac{2428}{81}+\frac{67}{9}\zeta_2-\frac{22}{9}\zeta_3-12\zeta_2^2\right]\nonumber\\
&&+C_F
N_F\left[-\frac{328}{81}-\frac{10}{9}\zeta_2+\frac{4}{9}\zeta_3\right],
\end{eqnarray}
where ${\L}=\ln\frac{\mu^2\overline N^2}{Q^2}$.  And finally, for
the Higgs case, we have
\begin{eqnarray}
{\cal
M}_{N,g}^{(1)}&=&C_A\left[2
{\L}^2+2\zeta_2\right],\nonumber\\
{\cal
M}_{N,g}^{(2)}&=&C_A^2\left({1\over
2}\right)[2
{\L}^2+2\zeta_2]^2+C_AC_A\left[\frac{22}{9}{\L}^3+\left(\frac{134}{9}-4\zeta_2\right)
{\L}^2+\left(\frac{808}{27}-28\zeta_3\right)
{\L}\right]\nonumber\\
&&-C_AN_F\left[
\frac{4}{9}{\L}^3+\frac{20}{9}
{\L}^2+\frac{112}{27}{\L}\right]\nonumber\\
&&+
C_AC_A\left[\frac{2428}{81}+\frac{67}{9}\zeta_2-\frac{22}{9}\zeta_3-12\zeta_2^2\right]\nonumber\\
&&+C_A
N_F\left[-\frac{328}{81}-\frac{10}{9}\zeta_2+\frac{4}{9}\zeta_3\right].
\end{eqnarray}
For all three processes, we have $G_N^{(0)}=1$.

From the above results it is clear that the logarithms $\rm L$ and
$\L$ vanish when we set: $\mu^2=\mu^2_I\equiv Q^2/{\overline N}^p$.
Of course, this has to be the case as the matching coefficients
should be logarithmically free, and we can write
\begin{eqnarray}
{\cal
M}_{N}\left(\frac{Q^2}{\mu^2},{\overline
N},\alpha_s(\mu^2)\right)={\cal
M}_{N}\left(\ln\left(\frac{Q^2}{{\overline
N}^p\mu^2}\right),\alpha_s(\mu^2)\right),
\end{eqnarray}
and
for $\mu^2=\mu_I^2\equiv \frac{Q^2}{{\overline N}^p}$ we
have
\begin{equation}
{\cal M}_{N}\left(\ln\left(\frac{Q^2}{{\overline
N}^p\mu^2_I}\right),\alpha_s(\mu^2)\right)={\cal
M}_N(\alpha_s(\mu_I^2)).
\end{equation}
These observations are valid to all orders in perturbation
theory\cite{Man03} and they  lead to a strong constraint on the
anomalous dimensions of the effective operators on both sides of
the matching scale. Another interesting feature emerges from the
results of the DY and Higgs cases, ${\cal M}_{N,q}^{(i)}$ and
${\cal M}_{N,g}^{(i)}$, $i=1,2$: One can simply get the latter
from the former by replacing the overall factor $C_F$  with $C_A$
in the non-Abelian part. The Abelian part exponentiates and hence
all occurrence of $C_F$ shall be replaced by $C_A$. In this sense,
the matching coefficients seem to be universal.  This could be
argued based on that in the soft gluon limit, only the color
charges of annihilating quarks and gluons are relevant.

Following the same steps as we did after the first stage matching
at $Q^2$, we need now to consider the running of the effective
operators that were used to perform the matching at $\mu_I$.
However at and below the scale $\mu_I$ they are just the
conventional PDFs taken to the limit $z\rightarrow 1$. As such,
the running of the effective operators (the PDFs) is governed by
the well-known DGLAP (Dokshitzer-Gribov-Lipatov-Altarelli-Parizi)
evolution equation with anomalous dimension
\begin{eqnarray}
\gamma^N_{2,{(q,g)}}=A_{(q,g)}\ln {\overline N}^2-2B_{2,{(q,g)}},
\end{eqnarray}
where $A_{(q,g)}$ and $B_{2,{(q,g)}}$ are given in Eqs.~(\ref {a})
and (\ref {b}). We include the running effects in
\begin{eqnarray}
I_2=2\int_{\mu_I}^{\mu_F}\frac{d\mu}{\mu}\gamma_{2,{(q,g)}},
\end{eqnarray}
where $\mu_F$ is the factorization scale for parton distributions.

The resummed factorization coefficient functions for DY and Higgs
are
\begin{eqnarray}
\label{asd} G_{N,(q,g)}(Q) &=& \vert C_{(q,g)}(\alpha_s(Q))\vert^2
e^{I_1(Q,\mu_I)} \times {\cal M}_{N,(q,g)}(\alpha_s(\mu_I))
e^{I_2(\mu_I,\mu_F)},
\end{eqnarray}
where we have omitted $C_\phi^2$ for Higgs production. [The
definition of $I_1$ and $I_2$ differs by a minus sign from Ref.
\cite{IdiJi05}.] Anticipating the discussion of the next section,
we will set the factorization scale $\mu_F=Q$. The above equation
can be brought into an equivalent form by exploiting the running
of $\alpha_s$ from $\mu_I$ to $Q$ in ${\cal
M}_{N,(q,g)}(\alpha_s(\mu_I))$;
\begin{eqnarray}
{\cal M}_{N,(q,g)}(\alpha_s(\mu_I^2))={\cal
M}_{N,(q,g)}(\alpha_s(Q^2))~\exp[I_3],
\end{eqnarray}
where
\begin{eqnarray}
\label {BC} I_3=-2\int_{\mu_I}^{Q}\frac{d\mu}{\mu}\triangle
B_{(q,g)},
\end{eqnarray}
where
\begin{eqnarray}
\triangle B_{(q,g)}\equiv -\beta(\alpha_s)\frac{d\ln {\cal
M}_{N,(q,g)}}{d\ln \alpha_s}.
\end{eqnarray}
 The last two equations are also true for the DIS case. Thus
we write
\begin{equation}
\label{csr} G_N(Q) = {\cal F}(\alpha_s(Q)) e^{I(\lambda,
\alpha_s(Q))},
\end{equation}
where ${\cal F} = \vert C_{(q,g)}(\alpha_s(Q))\vert ^2{\cal
M}_{(q,g)}(\alpha_s(Q))$ depends only on $\alpha_s(Q)$. The
subscript $N$ of ${\cal M}$ has been omitted since there is not
any large logarithmic dependence in the matching coefficients.
$I=I_1+I_2+I_3$ is a function of $\lambda=\beta_0\ln \overline N
\alpha_s(Q)$ and $\alpha_s(Q)$ with all leading and sub-leading
large logarithms resummed.

Since the cross section $\sigma_N$ in Eq.~(2) is independent of
the intermediate scale $\mu_I$, then from Eq.~(\ref{asd}) and the
definitions of $\gamma_1$ and $\gamma_2$ we get the following
relation for DY and Higgs;
\begin{eqnarray}
\frac{d\ln {\cal
M}_{N,(q,g)}(\alpha_s(\mu^2),\L)}{d\ln
\mu}=\left[2\gamma_2-2\gamma_1\right]_{(q,g)}=2[A\L+f]_{(q,g)},
\end{eqnarray}
from which we get
\begin{eqnarray}
\frac{d\ln {\cal
M}_{N,(q,g)}(\alpha_s(\mu^2),\L)}{d\ln
\mu}\Big
|_{\mu=\mu_I}=2f_{(q,g)}(\alpha_s(\mu_I^2)),~~~~\mu_I=\frac{Q}{\overline
N}
\end{eqnarray}
where
$A_{(q,g)}$ are given in Eq.~(\ref {a}) and $f_{(q,g)}$ are given in
Eq.~(\ref {f}). The last equation sheds light on the physical
meaning of the functions $f_{(q,g)}$: It is the anomalous dimension
of the matching coefficient ${\cal M}$ evaluated at the intermediate
scale $\mu_I$. Here we see that the universality of these functions
could be explained by the fact that ${\cal M}_{(q,g)}$ are
themselves universal.

The last equation also shows the same $A_{(q,g)}$ appears in the
logarithmic parts of $\gamma_{1,(q,g)}$ and $\gamma_{2,(q,g)}$,
because otherwise the logarithms at $\mu_I$ do not cancel in ${\cal
M}_N$.

For DIS a similar analysis is performed, however, we have to
consider only one-half of $I_2$ in Eq.~(\ref{asd}) since we match
onto a single PDF. With this we get
\begin{eqnarray}
\frac{d\ln {\cal
M}_{N,\rm DIS}(\alpha_s(\mu^2),{\rm L})}{d\ln
\mu}=\left[\gamma_2-2\gamma_1\right]_q=2[A{\rm L}+B_2+f]_q,
\end{eqnarray}
from which we obtain at the intermediate scale,
\begin{eqnarray}
\frac{d\ln {\cal M}_{N,{\rm
DIS}}(\alpha_s(\mu),{\rm L})}{d\ln
\mu}\Big
|_{\mu=\mu_I}=2[B_2+f]_q(\alpha_s(\mu_I^2)),~~~~\mu_I=\frac{Q}{\sqrt{\overline
N}}.
\end{eqnarray}
Here there is an extra contribution from $B_2$.

\section{Comparison With the Traditional Approach and Explicit Results To
N$^3$LL Order}

In this section we will illustrate the equivalence of the EFT
approach and the traditional one which relies on the refactorization
of hard processes as we mentioned in the introduction. The
renormalon problem in the later approach arises from doing
resummation uniformly for all moments, which will necessarily
encounter small scale $Q/N^p$ at fixed $Q$ when $N$ is sufficiently
large. The EFT approach avoids that by short-cutting the steps when
this scale becomes of order $\Lambda_{\rm QCD}$. We will start by
showing this first for DY and Higgs production, then will turn to
the DIS case. In the last subsection, we give the explicit form of
the relevant integrals obtained in the EFT approach.

\subsection{Drell-Yan and Higgs}

One of the well-known forms used to express the coefficient
function for DY and Higgs in moment space is the following
\cite{Managano}:
\begin{equation}
G_N(Q^2)=g_0(\alpha_s(Q^2))e^{I_\bigtriangleup}\Delta
C(\alpha_s(Q^2)),
\end{equation}
where we have normalized the Born term to $1$. The $g_0$ has a
conventional expansion form: $g_0=\sum_ia^ig_{0i}$. [In this
subsection, we omitted the subscript $q$ and $g$, intended for DY
and Higgs production.] The term $\bigtriangleup C$ has the only role
of cancelling the non-logarithmic contributions that appear in the
exponent. These contributions arise from the various $\zeta$-terms
in the Mellin transform of the ``plus'' distributions. The Sudakov
exponential term $I_\triangle$ is given by
\begin{eqnarray}
\label{tria}
I_{\triangle}=\int_0^1dz\frac{z^{N-1}-1}{1-z}\left[2\int_{Q^2}^{(1-z)^2Q^2}\frac{d\mu^2}{\mu^2}
A(\alpha_(\mu^2))
+D(\alpha_s((1-z)^2Q^2))\right],
\end{eqnarray}
where, as already mentioned, we set $\mu_F^2=Q^2$. As noted above,
$I_\Delta$ contains both a logarithmic and non-logarithmic
contribution. The quantities, $g_0$, $A$ and $D$ have the usual
expansion in $a_s$ and they are already known up to ${\cal
O}(\alpha_s^3)$ \cite{Vogt288}. The $A$ is identical to the
logarithmic coefficient in $\gamma_1$ and $\gamma_2$. It is our
aim to relate these quantities with those that appear in $G_N$ of
Eq.~(\ref {asd}). For this we follow the procedure outlined in
Appendices A, B and C of \cite{Catani:2003zt}. The integral in
$I_\triangle$ can be rewritten in terms of the already defined
$I_1$, $I_2$ and $I_3$,
\begin{eqnarray}
I\equiv I_1+I_2+I_3=I_{\triangle}+\ln
\triangle C(\alpha_s(Q^2)),
\end{eqnarray}
where the coefficient function
$\bigtriangleup C$ does not depend on
$\mu_I\sim Q/{\overline N}$.

To prove the above relation, we first use following expansion;
\begin{equation}
\label{z}
z^{N-1}-1=-\tilde{\Gamma}\left(1-\frac{\partial}{\partial\ln
\overline
N}\right)\theta\left(1-z-\frac{1}{\overline
N}\right)+{\cal O}(1/{\overline
N}),
\end{equation}
where the $\tilde \Gamma$ function is related to the usual gamma
function,
\begin{equation}
\tilde\Gamma\left(1-\frac{\partial}{\partial\ln\overline
N}\right)=1-\Gamma_2\left(\frac{\partial}{\partial\ln\overline
N}\right)\left(\frac{\partial}{\partial\ln\overline
N}\right)^2,
\end{equation}
where the first parenthesis in the right-hand side is the argument
of the $\Gamma_2$ function, and
\begin{eqnarray}
\Gamma_2(\epsilon)={1\over
\epsilon^2}[1-e^{-\gamma_E\epsilon}\Gamma(1-\epsilon)]=
-\frac{1}{2}\zeta_2-\frac{1}{3}\zeta_3\epsilon-{9\over
40}\zeta_2^2\epsilon^2+{\cal O}(\epsilon^3).
\end{eqnarray}
In Eq.~(\ref{z}) we used $(\partial/\partial\ln N) f(\ln {\overline
N})= (\partial/\partial\ln {\overline N})f(\ln {\overline N})$ for
an arbitrary function $f$. After some algebra,
$I_{\triangle}$ can be expressed
as
\begin{eqnarray}
I_{\triangle}&=&-\tilde{\Gamma}\left(1-\frac{\partial}{\partial\ln\overline
N}\right)\left\{\int_{Q^2/{\overline
N}^2}^{Q^2}\frac{d\mu^2}{\mu^2}\left[A(\alpha_s(\mu^2))\ln\frac{Q^2}{\mu^2}+\frac{1}{2}D(\mu^2)\right]
\right.\nonumber\\
&&+\left.\int_{Q^2}^{Q^2/{\overline
N}^2}\frac{d\mu^2}{\mu^2}A(\alpha_s(\mu^2))\ln\overline{N}^2\right\}.
\end{eqnarray}
The double derivative from $\tilde \Gamma$ acting on the curly
bracket above gives a contribution
\begin{eqnarray}
\Gamma_2\left(\frac{\partial}{\partial \ln
{\overline
N}}\right)\left[\frac{\partial}{\partial \ln
{\overline
N}}D(\alpha_s(Q^2/{\overline
N}^2))-4A(\alpha_s(Q^2/{\overline
N}^2))\right].
\end{eqnarray}
To compare $I_\Delta$ with the exponent $I=I_1+I_2+I_3$, we express
the latter in the form
\begin{eqnarray}
I_1+I_2+I_3&=&-\left\{\int_{Q^2/{\overline
N}^2}^{Q^2}\frac{d\mu^2}{\mu^2}\left[A(\alpha_s(\mu^2))\ln\frac{Q^2}{\mu^2}
+(B_1+{\bigtriangleup
B}+2B_2)\right]\right.\nonumber\\
&&\left.+\int_{Q^2}^{Q^2/{\overline
N}^2}\frac{d\mu^2}{\mu^2}A(\alpha_s(\mu^2))\ln{\overline
N}^2\right\},
\label{i-relation}
\end{eqnarray}
Matching the two integrals, we get
\begin{eqnarray}
\label{gamma} &-&\int_{Q^2/{\overline
N}^2}^{Q^2}\frac{d\mu^2}{\mu^2}(B_1+\bigtriangleup
B+2B_2)(\alpha_s(\mu^2))\nonumber \\~~~~ &=&
\Gamma_2\left(\frac{\partial}{\partial \ln {\overline
N}}\right)\left[\frac{\partial}{\partial \ln {\overline
N}}D(\alpha_s(Q^2/{\overline N}^2))-4A(\alpha_s(Q^2/{\overline
N}^2))\right]\nonumber
\\
&& ~~~-\frac{1}{2}\int_{Q^2/{\overline
N}^2}^{Q^2}\frac{d\mu^2}{\mu^2}D(\alpha_s(\mu^2))+\ln \bigtriangleup
C(\alpha_s(Q^2)).
\end{eqnarray}
The above equation can be solved by perturbative expansion in
$\alpha_s$.

If the equality given in Eq.~(\ref{gamma}) holds to all values of
${\overline N}$, then for ${\overline N}=1$ we get
\begin{equation}
\ln \bigtriangleup
C(\alpha_s(Q^2))=-\Gamma_2(\partial_{\alpha_s})\left[\partial
_{\alpha_s}D(\alpha_s(Q^2/{\overline
N}^2))-4A(\alpha_s(Q^2/{\overline N}^2))\right]\Bigg|_{{\overline
N}=1},
\end{equation}
where we follow \cite{Catani:2003zt} and replace the derivative
$\partial/\partial \ln {\overline N}$ with $\partial_{\alpha_s}$
where
\begin{equation}
\partial_{\alpha_s}\equiv
2\frac{d\alpha_s(\mu^2)}{d\ln\mu^2}\frac{\partial}{\partial\alpha_s}=-2\beta(\alpha_s)
\alpha_s\frac{\partial}{\partial\alpha_s}.
\end{equation}
and, hence,
\begin{equation}
\left(\frac{\partial}{\partial \ln
{\overline
N}}\right)f(\alpha_s(Q^2/{\overline
N}^2))=\partial_{\alpha_s}f(\alpha_s(Q^2/{\overline
N}^2)),
\end{equation}
where $f$ is an arbitrary function.

Applying one more $\partial/\partial \ln {\overline
N}=\partial_{\alpha_s}$ on both sides of Eq.~(\ref {gamma}) we get
our master relation
\begin{equation}
\label{MAS} 2( B_1+\triangle
B+2B_2)(\alpha_s(\mu^2))=D(\alpha_s(\mu^2))+\partial_{\alpha_s}\Gamma_2(\partial_{\alpha_s})\left[4A-\partial_{\alpha_s}
D\right](\alpha_s(\mu^2)).
\end{equation}
which can easily be solved for $D^{(i)}$ order by order in
$\alpha_s$. As an example, let us expand both sides up to ${\cal
O}(\alpha_s^4)$. First, we work out the expansion of the
$\bigtriangleup B$ term. From Eq.~(\ref {BC}), we get
\begin{eqnarray}
\label{b00}
\triangle B^{(0)}_{(q,g)}&=&\triangle
B^{(1)}_{(q,g)}=0,\nonumber\\
\bigtriangleup
B^{(2)}_{(q,g)}&=&-\beta_0M_{N,(q,g)}^{(1)},\nonumber\\
\bigtriangleup
B^{(3)}_{(q,g)}&=&-\beta_0\left[2M_{N,(q,g)}^{(2)}-\left({\cal
M}_{N,(q,g)}^{(1)}\right)^2\right]-\beta_1{\cal
M}_{N,(q,g)}^{(1)}, \\
\triangle B^{(4)}_{(q,g)}&=&-\beta_0\left[3{\cal
M}_{N,(q,g)}^{(3)}-3{\cal M}_{N,(q,g)}^{(1)}{\cal
M}_{N(q,g)}^{(2)}+\left ({\cal
M}_{N,(q,g)}^{(1)}\right)^3\right]\nonumber \\
&& -\beta_1\left[2{\cal M}_{N,(q,g)}^{(2)}-\left({\cal
M}_{N,(q,g)}^{(1)}\right)^2\right]\nonumber\\
&&-\beta_2{\cal
M}_{N,(q,g)}^{(1)}.
\end{eqnarray}
Noticing that
$B_{1,(q,g)}^{(i)}+2B_{2,(q,g)}^{(i)}=-f^{(i)}_{(q,g)}$ and using
the expansion of $\Gamma_2$, we get $D^{(i)}$
\begin{eqnarray}
\label{d00}
D^{(0)}_{(q,g)}&=&D^{(1)}_{(q,g)}=0,\nonumber\\
D^{(2)}_{(q,g)}&=&-2f_{(q,g)}^{(2)}+2\triangle B^{(2)}_{(q,g)}
+4\beta_0\zeta_2 A^{(1)}_{(q,g)},
\nonumber\\
D^{(3)}_{(q,g)}&=&-2f^{(3)}_{(q,g)}+2\triangle
B_{(q,g)}^{(3)}+4\zeta_2\beta_1A^{(1)}_{(q,g)}+8\zeta_2\beta_0A^{(2)}_{(q,g)}+{32\over
3}\zeta_3\beta_0^2A^{(1)}_{(q,g)},
\nonumber\\
D^{(4)}_{(q,g)}&=&-2f^{(4)}_{(q,g)}+2\triangle
B_{(q,g)}^{(4)}+12\zeta_2\beta_0A^{(3)}_{(q,g)}+8\zeta_2\beta_1A^{(2)}_{(q,g)}+32\zeta_3\beta_0^2A^{(2)}_{(q,g)}\nonumber\\
&&+{80\over 3}\zeta_3\beta_0\beta_1A^{(1)}_{(q,g)}+{216\over
5}\zeta_2^2\beta_0^3
A^{(1)}_{(q,g)}-12\zeta_2\beta_0^2D^{(2)}_{(q,g)}.
\end{eqnarray}
Thus, apart from the coupling-constant running effects, $D$ is
essentially $-2f =2B_1+4B_2$.

From the last two equations we see that in order to get $D^{(k)}$,
the only same order information needed is $f^{(k)}$. All the
quantities needed to calculate $D^{(2)}$ and $D^{(3)}$ are known and
we get
\begin{eqnarray}
D^{(2)}_{(q,g)}&=&C_{(q,g)}\left\{C_A\left(-\frac{101}{27}+\frac{11}{3}\zeta_2+\frac{7}{2}\zeta_3\right)+
N_F\left(\frac{14}{27}-\frac{2}{3}\zeta_2\right)\right\}.
\end{eqnarray}
\begin{eqnarray}
D^{(3)}_{(q,g)}&=&
C_{(q,g)}C_A^2\left[-\frac{594058}{729}+\frac{98224}{81}\zeta_2+\frac{40144}{27}\zeta_3
-\frac{2992}{15}\zeta_2^2-\frac{352}{3}\zeta_2\zeta_3-384\zeta_5\right]\nonumber\\
&&+C_{(q,g)}C_AN_F\left[\frac{125252}{729}-\frac{29392}{81}\zeta_2-\frac{2480}{9}\zeta_3
+\frac{736}{15}\zeta_2^2\right]\nonumber\\
&&+C_{(q,g)}C_FN_F\left[\frac{3422}{27}-32\zeta_2-
\frac{608}{9}\zeta_3-\frac{64}{5}\zeta_2^2\right]\nonumber\\
&&+C_{(q,g)}N_F^2\left[-\frac{3712}{729}+\frac{640}{27}\zeta_2+\frac{320}{27}\zeta_3\right],
\end{eqnarray}
where $C_{(q,g)}=C_F$ for the DY case and $C_A$ for the Higgs
case. The above results agree with the recent calculation in
\cite{Vogt265,Laenen284,RAVI}. The result for the Higgs production
has already been reported on in \cite{JiPRL}.

The non-logarithmic contribution ${\cal
F}_{(q,g)}(Q^2)=\sum_ia^i{\cal F}^{(i)}_{(q,g)}=\vert
C(Q^2)\vert^2{\cal M}_N(Q^2)$ can be calculated from the
already-known results for $C^{(i)}_{(q,g)}(Q^2)$ and ${\cal
M}_{N,(q,g)}^{(i)}(\alpha_s(Q^2))$, or we can simply read them
from the well-known results for $G^{i,{\rm (s+v)}}(Q^2)$ through
Eq.~(\ref{master1}) and Eq.~(\ref{s+v});
\begin{eqnarray}
{\cal
F}_q^{(1)}&=&16 C_F (\zeta_2-1),
\nonumber\\
{\cal
F}_q^{(2)}&=&C_F^2\left[\frac{511}{4}-198\zeta_2-60\zeta_3+\frac{552}{5}\zeta_2^2\right]\nonumber\\
&&+C_FC_A\left[-\frac{1535}{12}+\frac{376}{3}\zeta_2+\frac{604}{9}\zeta_3-\frac{92}{5}\zeta_2^2\right]\nonumber\\
&&+C_FN_F\left[\frac{127}{6}-\frac{64}{3}\zeta_2+\frac{8}{9}\zeta_3\right],
\end{eqnarray}
for DY lepton-pair production. For the Higgs case, we have
\begin{eqnarray}
{\cal F}_g^{(1)}&=&16\zeta_2C_A,
\nonumber\\
{\cal F}_g^{(2)}&=&C_A^2\left[93+{1072\over 9}\zeta_2-{308\over
9}\zeta_3+92\zeta_2^2\right]\nonumber\\
&&+C_AC_F\left[-\frac{1535}{12}+\frac{376}{3}\zeta_2+\frac{604}{9}\zeta_3-\frac{92}{5}\zeta_2^2\right]\nonumber\\
&&+C_AN_F\left[-\frac{80}{3}-\frac{160}{9}\zeta_2+\frac{88}{9}\zeta_3\right]+C_FN_F\left[-{67\over
3}+16\zeta_3\right].
\end{eqnarray}
The above results agree with the ${\rm g}_{01}$ and ${\rm g}_{02}$
in \cite{Vogt265}. The $\gamma_E$ terms in the results of
\cite{Vogt265} are due to the use of $N$ instead of ${\overline
N}$ as in our case. It is very simple to also reproduce these
terms. We also notice that their results for the $g_{0i}$ do not
include the contributions from the non-logarithmic terms in
$I_\triangle$.

\subsection{DIS}

For the DIS case there are essentially two major differences. The
first is that the $D$ term in $I_\triangle$ is zero to all orders in
$\alpha_s$ \cite{ridolfi,gardi}. The second one comes from the ``jet
function'' which encodes the effects of collinear gluon emission
from the outgoing parton. So for DIS, the traditional approach
yields the following expression for the exponent in the coefficient
function $G_N(Q^2)$,
\begin{eqnarray}
I_{\rm
DIS}=\int_0^1dz\frac{z^{N-1}-1}{1-z}\left[\int_{Q^2}^{(1-z)Q^2}\frac{d\mu^2}{\mu^2}A_q(\alpha_s(\mu^2))+{\cal
B}_q(\alpha_s((1-z)Q^2))\right],
\end{eqnarray}
where again we set $\mu_F^2=Q^2$. We have used ${\cal B}$ here so
that it will not be confused with $B_i$'s introduced earlier.

We now follow the same procedure as for the DY case, rewriting
\begin{eqnarray}
I_{\rm DIS
}&=&-\tilde{\Gamma}\left(1-\frac{\partial}{\partial\ln\overline
N}\right)\left\{\int_{Q^2/{\overline
N}}^{Q^2}\frac{d\mu^2}{\mu^2}\left[A_q(\alpha_s(\mu^2))\ln\frac{Q^2}{\mu^2}+{\cal
B}_q(\mu^2)\right]
\right.\nonumber\\
&&+\left.\int_{Q^2}^{Q^2/{\overline
N}}\frac{d\mu^2}{\mu^2}A_q(\alpha_s(\mu^2))\ln\overline{N}\right\},
\end{eqnarray}
On the other hand, our result for DIS reads
\begin{eqnarray}
I_1+I_2+I_3&=&-\left\{\int_{Q^2/{\overline
N}}^{Q^2}\frac{d\mu^2}{\mu^2}\left[A_q(\alpha_s(\mu^2))\ln\frac{Q^2}{\mu^2}
+(B_{1,q}+{\bigtriangleup
B}_{\rm DIS}+B_{2,q})\right]\right.\nonumber\\
&&\left.+\int_{Q^2}^{Q^2/{\overline
N}}\frac{d\mu^2}{\mu^2}A_q(\alpha_s(\mu^2))\ln{\overline N}\right\}.
\label{i-relation}
\end{eqnarray}
Matching the two results above, and noting that
\begin{equation}
\left(\frac{\partial}{\partial \ln {\overline
N}}\right)f(\alpha_s(Q^2/{\overline N}))={1\over
2}\partial_{\alpha_s}f(\alpha_s(Q^2/{\overline N})),
\end{equation}
we get the final relation between EFT and traditional approaches
for the DIS case;
\begin{eqnarray}
&& (B_{1,q}+\bigtriangleup B_{\rm
DIS}+B_{2,q})(\alpha_s(\mu^2))\nonumber \\
&=&{\cal B}_q(\alpha_s(\mu^2))+{1\over
2}\partial_{\alpha_s}\Gamma_2\left({1\over
2}\partial_{\alpha_s}\right)\left[A_q-{1\over
2}\partial_{\alpha_s}{\cal B}_q\right](\alpha_s(\mu^2)),
\end{eqnarray}
from which we can solve for ${\cal B}_q^{(i)}$. Up to third order we
have
\begin{eqnarray}
{\cal B}^{(1)}_q &=& -B_{2,q}^{(1)}, \nonumber \\
{\cal B}^{(2)}_q &=& -B_{2,q}^{(2)} - f_q^{(2)} + \Delta B_{\rm DIS}^{(2)} +
\frac{1}{2}
\zeta_2\beta_0 A_q^{(1)},  \nonumber \\
  {\cal B}^{(3)}_q &=& -B_{2,q}^{(3)} - f_q^{(3)} + \Delta B_{\rm DIS}^{(3)}
+\beta_0\zeta_2A^{(2)}_q
  + \frac{1}{2}\zeta_2\beta_1 A_q^{(1)} +  \frac{2}{3}\zeta_3\beta_0^2
A_q^{(1)}.
\end{eqnarray}
Therefore, apart from running effects, ${\cal B}_q$ is essentially
$-B_{2,q}-f_q$. More explicitly, we get
\begin{eqnarray}
{\cal B}^{(1)}_q&=& -3C_F, \nonumber
\\
{\cal B}^{(2)}_q&=&
C_F^2\left[-\frac{3}{2}+12\zeta_2-24\zeta_3\right]+C_FC_A\left[-\frac{3155}{54}+\frac{44}{3}\zeta_2+40\zeta_3\right]\nonumber
\\
&&+C_FN_F\left[\frac{247}{27}-\frac{8}{3}\zeta_2\right],\nonumber\\
{\cal B}^{(3)}_q&=&
C_F^3\left[-\frac{29}{2}-18\zeta_2-68\zeta_3-\frac{288}{5}\zeta_2^2+32\zeta_2\zeta_3+240\zeta_5\right]\nonumber\\
&&+C_AC_F^2\left[-46+287\zeta_2-\frac{712}{3}\zeta_3-\frac{272}{5}\zeta_2^2-16\zeta_2\zeta_3-120\zeta_5\right]\nonumber\\
&&+C_A^2C_F\left[-\frac{599375}{729}+\frac{32126}{81}\zeta_2+\frac{21032}{27}\zeta_3-\frac{652}{15}\zeta_2^2-\frac{176}{3}\zeta_2\zeta_3-232\zeta_5\right]\nonumber\\
&&+C_F^2N_F\left[\frac{5501}{54}-50\zeta_2+\frac{32}{9}\zeta_3\right]+C_FN_F^2\left[-\frac{8714}{729}+\frac{232}{27}\zeta_2-\frac{32}{27}\zeta_3\right]
\nonumber\\
&&+C_AC_FN_F\left[\frac{160906}{729}-\frac{9920}{81}\zeta_2-\frac{776}{9}\zeta_3+\frac{208}{15}\zeta_2^2\right].
\end{eqnarray}
Those results agree with the ones in Ref.~\cite{Vogt288}. Similar
to the case of DY and Higgs, we get after simple calculation
\begin{eqnarray}
{\cal F}_{\rm DIS}^{(1)}&=&16 C_F (-9-2\zeta_2),
\nonumber\\
{\cal
F}_{\rm DIS}^{(2)}&=&C_F^2\left[\frac{331}{8}+{111\over
2}\zeta_2-66\zeta_3+\frac{4}{5}\zeta_2^2\right]\nonumber\\
&&+C_FC_A\left[-\frac{5465}{72}-\frac{1139}{18}\zeta_2+\frac{464}{9}\zeta_3+\frac{51}{5}\zeta_2^2\right]\nonumber\\
&&+C_FN_F\left[\frac{457}{36}+\frac{85}{9}\zeta_2+\frac{4}{9}\zeta_3\right].
\end{eqnarray}
Again
these results agree with $g_{01}^{\rm DIS}$ and
$g_{02}^{\rm
DIS}$.

\subsection{Drell-Yan Coefficient Function Using DIS Parton Distributions}

If one calculates the Drell-Yan coefficient function in terms of the
DIS parton distributions, one has
\begin{eqnarray}
   \Delta_N &=& G_{N, q}/G_{N, \rm DIS}^2 \nonumber \\
        &\sim &
\int_0^1dz\frac{z^{N-1}-1}{1-z}\left[2\int_{(1-z)Q^2}^{(1-z)^2Q^2}\frac{d\mu^2}{\mu^2}
A_q(\alpha_(\mu^2)) \right.\nonumber
\\&& \left.+D_q(\alpha_s((1-z)^2Q^2)-2{\cal B}_q((1-z)Q^2 )\Large\right],
\end{eqnarray}
We have seen from the last two subsections that if one ignores the
running effects, $D_q\sim 2B_1+4B_2$ and ${\cal B}_q \sim B_1+B_2$.
Hence the last two terms in the above equation is just $\sim 2B_2$
in EFT, negative of the coefficient in front of $\delta(1-x)$ in the
DGLAP splitting function.

\subsection{Performing the Integrals}

Another way to compare the EFT results with the traditional ones is
to carry out the integral $I_1+I_2+I_3$ directly, and compare the
final form of the resummed result. We wish also to show that the way
we arrive at the final result is much simpler than the existing one
in the literature.

Specializing for the DY and Higgs case, the integral is then,
\begin{eqnarray}
\label {simple I} I_1+I_2+I_3&=&\int_{Q^2/{\overline
N}^2}^{Q^2}\frac{d\mu^2}{\mu^2}\left[A_{(q,g)}(\alpha_s(\mu^2))\ln\frac{\mu^2{\overline
N}^2}{Q^2} -({\bigtriangleup B}_{(q,g)}-f_{(q,g)})\right].
\end{eqnarray}
We also need the solution of the renormalization group equation for
$\alpha_s(\mu^2)$. Adopting the notation of
Ref.~\cite{Catani:2003zt} we have
\begin{eqnarray}
\alpha_s(\mu^2)&=&\frac{\alpha_s(Q^2)}{l}\left\{1-\frac{\alpha_s(Q^2)}{l}\frac{b_1}{b_o}\ln
l\right.\nonumber \\
&+&\left.\left(\frac{\alpha_s(Q^2)}{l}\right)^2\left[\frac{b_1^2}{b_0^2}(\ln^2
l-\ln l +l-1)-\frac{b_2}{b_0}(\ln l-1)\right] +{\cal
O}(\alpha_s(Q^2))\right\},
\end{eqnarray}
where $l=1+b_0\alpha_s(Q^2)\ln \mu^2/Q^2$ and
$b_i=\frac{1}{(4\pi)^{i+1}}\beta_i$. Let us start with the
contribution of the  $A^{(1)}_{(q,g)}$ term. Changing the
integration variable from $\mu^2$ to $l$, this contribution gives
\begin{eqnarray}
\label{A_1}
I_{A_1}&=&\frac{A^{(1)}_{(q,g)}}{4\pi
b_0}\int_{1-2\lambda}^{1}\frac{dl}{l}\left\{1-\alpha_s(Q^2)\frac{b_1}{b_0}\frac{\ln
l}{l}
+\left(\frac{\alpha_s(Q^2)}{l}\right)^2\left[\frac{b_1^2}{b_0^2}\left[\ln^2
l-\ln
l+l-1\right]\right.\right.\nonumber\\
&&\left.\left.-\frac{b_2}{b_0}\left(\ln
l-1\right)\right]\right\}\left(2\ln
{\overline
N}+\frac{l-1}{b_0\alpha_s(Q^2)}\right),
\end{eqnarray}
where $\lambda\equiv b_0\alpha_s(Q^2)\ln {\overline N}$. The last
equation includes a pattern that repeats itself when other
contributions are included. Taking as a working rule that $\ln
{\overline N}\sim (1/\alpha_s(Q^2))$, the last two terms give rise
to comparable contributions, however inside the curly brackets we
have expansion in $\alpha_s(Q^2)$. Thus the hierarchy is manifest.
Carrying out the integrals in Eq.~(\ref{A_1}) is very simple and
we get
\begin{eqnarray}
I_{A_1}&=&\ln {\overline N}\left\{\frac{A^{(1)}_{(q,g)}}{4\pi
b_0}\left[\frac{2\lambda+(1-2\lambda)\ln
(1-2\lambda)}{\lambda}\right]\right\}\nonumber
\\ &+& \frac{A^{(1)}_{(q,g)}b_1}{4\pi b_0^3}\left[2\lambda+\ln
(1-2\lambda)+{1\over
2}\ln^2(1-2\lambda)\right]\nonumber\\
&&+\alpha_s(Q^2)\frac{A^{(1)}_{(q,g)}b_1^2}{4\pi
b_o^4}\left[2\lambda^2+2\lambda\ln
(1-2\lambda)+{1\over
2}\ln^2(1-2\lambda)\right]\frac{1}{1-2\lambda}.
\end{eqnarray}
Expanding the $\lambda$-terms in the last equation, we get a sum
of the form $\alpha_s^n(Q^2)\ln^{n+1} {\overline N}$ from the
first term, $\alpha_s^n(Q^2)\ln^n {\overline N}$ from the second
term, and $\alpha_s^{n+1}\ln^n {\overline N}$ from the last term.
These are commonly called: leading logarithms (LL),
next-to-leading logarithms (NLL) and next-to-next-to leading
logarithms (NNLL), respectively. Higher logarithmic accuracies
follow easily.

 Consider now the
 contribution from $A^{(2)}_{(q,g)}$.
 Similar to the $A^{(1)}$ contribution we
 get
 \begin{eqnarray}
 I_{A_2}&=&\frac{A^{(2)}_{(q,g)}}{(4\pi)^2b_0}\int_{1-2\lambda}^{1}\frac{dl}{l^2}\alpha_s(Q^2)\left[1-2\alpha_s(Q^2)\frac   {b_1}{b_o}\frac{\ln
  l}{l}+{\cal O}(\alpha_s^3(Q^2))\right]\nonumber \\ && ~~\times
 \left(2\ln {\overline N}+\frac{l-1}{b_0\alpha_s(Q^2)}\right),
 \end{eqnarray}
 so we see that
 $A^{(2)}_{(q,g)}$ does not contribute  to the LL
 but starts from NLL. This
 contribution is
 \begin{eqnarray}
 I_{A_2}
 &=&-\frac{A^{(2)}_{(q,g)}}{(4\pi)^2b_0^2}\left[2\lambda+\ln
 (1-2\lambda)\right]-\alpha_s(Q^2)\frac
 {A^{(2)}_{(q,g)}b_1}{(4\pi)^2b_0^3}\left[2\lambda
+ 2\lambda^2+\ln
 (1-2\lambda)\right].
 \end{eqnarray}
 From the  $A^{(3)}_{(q,g)}$ term we get
 \begin{eqnarray}
 I_{A_3}
 &=&\frac{A^{(3)}_{(q,g)}}{(4\pi)^3b_0}\int_{1-2\lambda}^{1}\frac{dl}{l^3}\alpha_s^2(Q^2)\left[1+{\cal
 O}(\alpha_s(Q^2))\right]\left(2\ln
 {\overline
 N}+\frac{l-1}{b_0\alpha_s(Q^2)}\right),
 \end{eqnarray}
 which is a NNLL contribution;
 \begin{eqnarray}
 I_{A_3}&=&\alpha_s(Q^2)\frac{A^{(3)}_{(q,g)}}{(4\pi)^3b_0^2}~\frac{2\lambda}{1-2\lambda}.
 \end{eqnarray}

 The contribution from the term $\triangle B^{(i)}-f^{(i)}$ starts
 at NNLL accuracy since this term vanishes for $i=0,1$. From
 Eq.~(\ref {d00}) we have $\triangle
 B^{(2)}_{(q,g)}-f^{(2)}_{(q,g)}=(1/2)(D^{(2)}-4\beta_0\zeta_2A^{(1)}_{(q,g)})$.
 The contribution of this term gives
 \begin{eqnarray}
 I_{B_2}&=&-\frac{1}{(4\pi^2)}\frac{1}{b_0\alpha_s(Q^2)}[\triangle
 B^{(2)}_{(q,g)}-f^{(2)}_{(q,g)}]\int_{1-2\lambda}^{1}\frac{dl}{l^2}\alpha_s^2(Q^2),
 \end{eqnarray}
 which is a NNLL contribution;
 \begin{eqnarray}
 I_{B_2}&=&\alpha_s(Q^2)\frac{1}{(4\pi)^2b_0}\left[4\beta_0\zeta_2A^{(1)}_{(q,g)}-D^{(2)}_{(q,g)}\right]~\frac{\lambda}  {1- \lambda}.
 \end{eqnarray}

 Writing the sum of all contributions already obtained in the form
 of
 \begin{eqnarray}
 I_{A_1} + I_{A_2} + I_{A_3} + I_{B_2} = \ln {\overline
 N}g^{(1)}_{(q,g)}+g^{(2)}_{(q,g)}+\alpha_s(Q^2)g^{(3)}_{(q,g)},
 \end{eqnarray}
 we
 get
 \begin{eqnarray}
 g^{(1)}_{(q,g)}(\lambda)&=&\frac{A^{(1)}_{(q,g)}}{4\pi
 b_0}\left[\frac{2\lambda+(1-2\lambda)\ln
 (1-2\lambda)}{\lambda}\right],\nonumber\\
 g^{(2)}_{(q,g)}(\lambda)&=&-\frac
 {A^{(2)}_{(q,g)}}{(4\pi)^2b_0^2}\left[2\lambda+\ln(1-2\lambda)\right]+\frac{A^{(1)}_{(q,g)}b_1}{4\pi
 b_0^3}\left[2\lambda+\ln(1-2\lambda)+\frac{1}{2}\ln^2(1-2\lambda)\right],\nonumber
 \\
 g^{(3)}_{(q,g)}(\lambda)&=&\left[\frac{4\zeta_2A^{(1)}_{(q,g)}}{4\pi}-\frac{D^{(2)}_{(q,g)}}{(4\pi)^2b_0}\right]~\frac {\lambda}{1-2\lambda}
+\frac{A^{(1)}_{(q,g)}b_1^2}{4\pi
b_0^3}\left[2\lambda+2\lambda\ln
(1-2\lambda)+{1\over 2}\ln^2
(1-2\lambda)\right]\nonumber\\
&&+\frac{A^{(1)}_{(q,g)}b_2}{4\pi
b_0^3}\left[2\lambda+\ln
(1-2\lambda)+\frac{2\lambda^2}{1-2\lambda}\right]+\frac{2A^{(3)}_{(q,g)}}{(4\pi)^3b_0^2}~\frac{\lambda^2}{1-2\lambda}\nonumber\\
&&-\frac{A^{(2)}_{(q,g)}b_1}{(4\pi)^2b_0^3}\left[2\lambda+2\lambda^2+\ln
(1-2\lambda)\right]~\frac{1}{1-2\lambda}.
\end{eqnarray}
 The above functions sum the large logarithms to LL, NLL and ${\rm
 N}^2$LL, respectively. It is straightforward to get also the
 $\alpha_s^2g^{(4)}$ which resumms the ${\rm N}^3$LL. It will
 contain contributions from $A^{(i)}_{(q,g)}$ up to $i=4$ and from
 $D^{(2)}_{(q,g)}$ and $D^{(3)}_{(q,g)}$. The yet uncalculated
 quantity $A^{(4)}_{(q,g)}$ is the only missing piece to complete
 the ${\rm N}^3$LL resummation program. The above results for
 $g^{(i)}$ agree with those in \cite{Catani:2003zt,Vogt146}. We
 remind the reader that we have set the factorization scale and the
 renormalization scale equal to $Q^2$ and the $\gamma_E$ dependence
 is hidden in $\overline N$ used throughout. The analysis for the
 DIS case can be performed similarly and one also gets agreement
 with the known results.

 \section{Conclusion}

 Threshold resummation of logarithmic enhancements due to soft
 gluon radiation has been performed using the methodology of
 effective field theory. This method works to any desired
 (subleading) logarithmic accuracy and it is completely equivalent
 to the more conventional, factorization-based techniques. This has
 been illustrated to all three inclusive processes we considered:
 DIS, DY and the SM Higgs production.

 Conceptually and technically, however, this approach is much less
 complicated and it is physically more transparent than other ones.
 Working perturbatively in moment space (and for large values of
 $N$) we found that one does \emph{not} need to introduce any
 additional nonperturbative quantities (other than the conventional
 PDFs), as is usually the case in the traditional approaches. All
 the quantities needed to get the resummed coefficient functions
 are straightforwardly obtained from fixed-order calculations of
 the form factors (which supply the $C^{(i)}$ and the
 $\gamma_1^{(i)}$), the Altarelli-Parisi splitting kernels (which
 supply the $\gamma_2^{(i)}$) and the cross section for real gluon
 emission in the soft limit (from which we get the ${\cal
 M}^{(i)}$). It should be mentioned that the given treatment of DIS
 is applicable only in the Bjorken limit where one takes $Q^2$ to
 infinity first. However, for finite (but large) values of $Q^2$
 where the scale $Q^2(1-x)^2$ would emerge, a different treatment
 is needed.

 The method discussed in this paper can be extended straightforwardly
 to other processes.

 \section*{ACKNOWLEDGMENTS}
 We thank J. P. Ma for collaboration at the early stage of this
 project. A. I. and X. J. was supported by the U. S. Department of
 Energy via grant DE-FG02-93ER-40762. X. J. is also supported by a
 grant from National Natural Science Foundation of China. F. Y.
 thanks Werner Vogelsang for useful discussions related to the
 subject of the present paper. He is grateful to RIKEN, Brookhaven
 National Laboratory and the U.S. Department of Energy (contract
 number DE-AC02-98CH10886) for providing the facilities essential for
 the completion of his work.

 {\bf Note:} While this paper is in writing, a paper by T. Becher and
 M. Neubert appeared in archive (hep-ph/0605050) which also uses EFT
 to resum the large logarithms for DIS. In their paper, the jet
 function is similar to the matching coefficient ${\cal M}$ here.

 \vfill\eject

\section*{APPENDIX} 

 In this appendix, we collect the coefficient functions for
 deep-inelastic scattering, Drell-Yan and the Higgs production
 (within the large top-quark mass effective theory) to ${\cal 
 O}(\alpha_s^2)$ in the soft limit of full QCD. They are used to
 extract the matching coefficients ${\cal M}$ in Eqs. (34-36). As we
 have remarked in the main paper, these results must be reproduced by
 calculations of an EFT in which only the soft and collinear degrees
 of freedom are taken into account. For DIS (see
 Refs.~\cite{Zil1,Zil2}) , Drell-Yan (see Refs.~\cite{Matsuura}) and
 Higg production (see Refs.~\cite{Catani01,Har01}), we have
 \begin{eqnarray}
 G^{(2),{\rm s+v}}_{\rm DIS}(x) &=&
 C_F^2\left\{\left[16{\cal
 D}_1(x)+12{\cal
 D}_0(x)+\delta(1-x)\left(\frac{9}{2}-8\zeta_2\right)\right]\ln^2\left(\frac{Q^2}{\mu^2}\right)\right.\nonumber\\
 &&+\Big[24{\cal
 D}_2(x)-12{\cal
 D}_1(x)-(45+32\zeta_2){\cal
 D}_0(x)\nonumber\\
 &&\left.+\delta(1-x)\left(-\frac{51}{2}-12\zeta_2+40\zeta_3\right)\right]\ln \left( \frac{Q^2}{\mu^2}\right)\nonumber\\
 &&\left.+8{\cal
 D}_3(x)-18{\cal D}_2(x)-(27+32\zeta_2){\cal
 D}_1(x)+2\times 48
 \times
 \left(-\frac{3}{4}\right)\zeta_3{\cal
 D}_0(x)\right.\nonumber\\
 &&\left.+\left(\frac{51}{2}+36\zeta_2+64\zeta_3\right){\cal
 D}_0(x)+\delta(1-x)\left(\frac{331}{8}+69\zeta_2-78\zeta_3+6\zeta_2^2\right)\right\}\nonumber
 \\
 &&+C_FN_F\left\{\left[{4\over 3}{\cal
 D}_0(x)+\delta(1-x)\right]\ln^2\left(
 \frac{Q^2}{\mu^2}\right)+\left[{8\over 3}{\cal D }_1(x)-{58\over
 9}{\cal D}_0(x)\right.\right.\nonumber
 \\
 &&\left.-\delta(1-x)\left({19\over 3}+{16\over
 3}\zeta_2\right)\right]\ln \left(\frac{Q^2}{\mu^2}\right) +{4\over
 3}{\cal D}_2(x)-{58\over 9}{\cal D}_1(x)+\left({247\over
 27}-{8\over 3}\zeta_2\right){\cal
 D}_0(x)\nonumber\\
 &&\left.+\delta(1-x)\left({457\over
 36}+{38\over
 3}\zeta_2+{4\over
 3}\zeta_3\right)\right\}\nonumber\\
 &&+C_AC_F\left\{\left[-{22\over
 3}{\cal
 D}_0(x)-{11\over
 2}\delta(1-x)\right]\ln^2 \left( \frac{Q^2}{\mu^2}\right)\right.\nonumber\\
 &&+\left[-{44\over
 3}{\cal D}_1(x)+\left({367\over
 9}-8\zeta_2\right){\cal
 D}_0(x)+\left({215\over
 6}+{88\over
 3}\zeta_2-12\zeta_3\right)\delta(1-x)\right]\ln \left( \frac{Q^2}{\mu^2}\right)\nonumber\\
 &&-{22\over 3}{\cal D}_2(x)+\left({367\over 9}-8\zeta_2\right){\cal
 D}_1(x)+36\zeta_3{\cal D}_0(x) \nonumber \\ && +\left(-{3155\over
 54}+{44\over 3}\zeta_2+4\zeta_3\right){\cal
 D}_0(x)\nonumber\\
 &&+\left.\delta(1-x)\left(-{5465\over
 72}-{251\over
 3}\zeta_2+{140\over 3}\zeta_3+{71\over
 5}\zeta_2^2\right)\right\}.
 \end{eqnarray}
 \begin{eqnarray}
 G^{(2),{\rm
 s+v}}_q(z) &=&C_F^2\left\{\left[64{\cal
 D}_1(z)+48{\cal
 D}_0(z)+\delta(1-z)(18-32\zeta_2)\right]\ln^2 \left(\frac{Q^2}{\mu^2}\right)\right.\nonumber\\
 &&+\left[192{\cal
 D}_2(z)+96{\cal
 D}_1(z)-(128+64\zeta_2){\cal
 D}_0(z)\right.\nonumber\\
 &&\left.+\delta(1-z)(-93+24\zeta_2+176\zeta_3)\right]\ln \left(\frac{Q^2}{\mu^2}\right)\nonumber\\
 &&+128{\cal
 D}_3(z)-(256+128\zeta_2){\cal
 D}_1(z)+256\zeta_3{\cal
 D}_0(z)\nonumber\\
 &&\left.+\delta(1-z)\left(\frac{511}{4}-70\zeta_2-60\zeta_3+{8\over 
 5}\zeta_2^2\right)\right\}\nonumber\\
 &&+C_FN_F\left\{\left[{8\over
 3}{\cal
 D}_0(z)+2\delta(1-z)\right]\ln^2\left(\frac{Q^2}{\mu^2}\right)\right.\nonumber\\
 &&\left.+\left[{32\over
 3}{\cal D}_1(z)-{80\over
 9}{\cal
 D}_0(z)-{34\over
 3}\delta(1-z)\right]\ln \left(\frac{Q^2}{\mu^2}\right)\right.\nonumber\\
 &&\left.+{32\over
 3}{\cal D}_2(z)-{160\over 9}{\cal
 D}_1(z)+\left({224\over 27}-{32\over
 3}\zeta_2\right){\cal
 D}_0(z)+\delta(1-z)\left({127\over
 6}-{112\over
 9}\zeta_2+8\zeta_3\right)\right\}\nonumber\\
 &&+C_AC_F\left\{\left(-{44\over
 3}{\cal
 D}_0(z)-11\delta(1-z)\right)\ln^2 \left(\frac{Q^2}{\mu^2}\right)\right.\nonumber\\
 &&+\left[-{176\over
 3}{\cal D}_1(z)+\left({536\over
 9}-16\zeta_2\right){\cal
 D}_0(z)+\left({193\over
 3}-24\zeta_3\right)\delta(1-z)\right]\ln \left(\frac{Q^2}{\mu^2}\right)\nonumber\\
 &&-{176\over
 3}{\cal D}_2(z)+\left({1072\over
 9}-32\zeta_2\right){\cal
 D}_1(z)+\left(-{1616\over 27}+{176\over
 3}\zeta_2+56\zeta_3\right){\cal
 D}_0(z)\nonumber\\
 &&\left.+\delta(1-z)\left(-{1535\over
 12}+{592\over
 9}\zeta_2+28\zeta_3-{12\over 5}\zeta_2^2\right)\right\}.
 \\
 G^{(2),{\rm s+v}}_g(z) &=&C_A^2\left\{\left[64{\cal D}_1(z)-{44\over
 3}{\cal D}_0(z)-32\zeta_2\delta(1-z)\right]\ln^2 \left(\frac{Q^2}{\mu^2}\right)\right.\nonumber\\
 &&+\left[192{\cal D}_2(z)-{176\over 3}{\cal
 D}_1(z)+\left({536\over
 9}-80\zeta_2\right){\cal
 D}_0(z)\right.\nonumber\\
 &&\left.+\delta(1-z)\left(-24-{88\over
 3}\zeta_2+152\zeta_3\right)\right]\ln \left(\frac{Q^2}{\mu^2}\right)\nonumber\\
 &&+128{\cal
 D}_3(z)-{176\over 3}{\cal D}_2(z)+\left({1072\over
 9}-160\zeta_2\right){\cal D}_1(z)\nonumber\\
 &&+\left(-{1616\over
 27}+{176\over 3}\zeta_2+312\zeta_3\right){\cal
 D}_0(z)\nonumber\\
 &&\left.+\delta(1-z)\left(93+{536\over 9}\zeta_2-{220\over
 3}\zeta_3-{4\over 5}\zeta_2^2\right)\right\}\nonumber\\
 &&+C_FN_F\delta(1-z)\left(4\ln\left(\frac{Q^2}{\mu^2}\right)-{67\over 3}+16\zeta_3\right)\nonumber\\
 &&+C_AN_F\left\{\left({8\over 3}{\cal
 D}_0(z)\right)\ln^2\left(\frac{Q^2}{\mu^2}\right)\right.\nonumber\\
 &&+\left[{32\over 3}{\cal D}_1(z)-{80\over 9}{\cal
 D}_0(z)+\delta(1-z)\left(8+{16\over 3}\zeta_2\right)\right]\ln
 \left(\frac{Q^2}{\mu^2}\right)
 \end{eqnarray}
 \begin{eqnarray}
 &&+{32\over 3}{\cal D}_2(z)-{160\over
 9}{\cal
 D}_1(z)+\left({224\over 27}-{32\over 3}\zeta_2\right){\cal
 D}_0(z)\nonumber\\
 &&\left.+\delta(1-z)\left(-{80\over 3}-{80\over
 9}\zeta_2-{8\over 3}\zeta_3\right)\right\}.
 \end{eqnarray}
 The Mellin transform of the above functions with respect to their
 arguments in the large ${\overline N}$ limit are,
 \begin{eqnarray}
 G^{(2),{\rm s+v}}_{N,\rm DIS}
 &=&C_F^2\left\{\left[8\mm-12\m+{9\over
 2}\right]\ln^2 \left(\frac{Q^2}{\mu^2}\right)\right.\nonumber\\
 &&+\left[-8\mmm-6\mm+(45+8\zeta_2)\m-{51\over
 2}-18\zeta_2+24\zeta_3\right]\ln \left(\frac{Q^2}{\mu^2}\right)
 \nonumber\\
 &&+2\mmmm+6\mmm-\left({27\over
 2}+4\zeta_2\right)\mm\nonumber\\
 &&\left.+\left(-{51\over
 2}-18\zeta_2+24\zeta_3\right)\m+{331\over
 8}+{111\over
 2}\zeta_2-66\zeta_3+{4\over
 5}\zeta_2^2\right\}\nonumber\\
 &&+C_FN_F\left\{\left[-{4\over
 3}\m+1\right]\ll+\left({4\over
 3}\mm+{58\over 9}\m-{19\over
 3}-4\zeta_2\right)\ln \left(\frac{Q^2}{\mu^2}\right)\right.\nonumber\\
 &&-{4\over 9}\mmm-{29\over
 9}\mm+\left(-{247\over 27}+{4\over
 3}\zeta_2\right)\m
 \nonumber\\ 
 &&\left.+{457\over 36}+{85\over
 9}\zeta_2+{4\over
 9}\zeta_3\right\}+C_AC_F\left\{\left[{22\over
 3}\m-{11\over
 2}\right]\ln^2\left(\frac{Q^2}{\mu^2}\right)\right.\nonumber\\
 &&+\left(-{22\over
 3}\mm-\left({367\over
 9}-8\zeta_2\right)\m+{215\over
 6}+22\zeta_2-12\zeta_3\right)\ln\left(\frac{Q^2}{\mu^2}\right)\nonumber\\
 &&+{22\over
 9}\mmm+\left({367\over
 18}-4\zeta_2\right)\mm+\left({3155\over
 54}-{22\over
 3}\zeta_2-40\zeta_3\right)\m\nonumber\\
 &&\left.-{5465\over
 72}-{1139\over 18}\zeta_2+{464\over
 9}\zeta_3+{51\over
 5}\zeta_2^2\right\}.\\
 G^{(2),{\rm s+v}}_{N,q}
 &=&C_F^2\left\{\left[32\mm-48\m+18\right]\ln^2 \left(\frac{Q^2}{\mu^2}\right)\right.\nonumber\\
 &&+\left[-64\mmm+48\mm+(128-128\zeta_2)\m-93+72\zeta_2+48\zeta_3\right]\ln \left(\frac{Q^2}{\mu^2}\right)\nonumber\\
 &&\left.+32\mmmm-(128-128\zeta_2)\mm+{511\over
 4}-198\zeta_2-60\zeta_3+{552\over
 5}\zeta_2^2\right\}\nonumber\\
 &&+C_FN_F\left\{\left[-{8\over
 3}\m+2\right]\ll+\left[{16\over
 3}\mm+{80\over 9}\m-{34\over
 3}+{16\over
 3}\zeta_2\right]\ln \left(\frac{Q^2}{\mu^2}\right)\right.\nonumber\\
 &&\left.-{32\over
 9}\mmm-{80\over 9}\mm-{224\over 27}\m+{127\over
 6}-{192\over
 9}\zeta_2+{8\over 9}\zeta_3\right\}
 \nonumber\\
 &&+C_FC_A\left\{\left[{44\over
 3}\m-11\right]\ll\right.\nonumber
 \end{eqnarray}
 \begin{eqnarray}
 &&+\left[-{88\over 3}\mm-\left({536\over
 9}-16\zeta_2\right)\m+{193\over 3}-{88\over
 3}\zeta_2-24\zeta_3\right]\ln\left(\frac{Q^2}{\mu^2}\right)\nonumber\\
 &&+{176\over
 9}\mmm+\left({536\over
 9}-16\zeta_2\right)\mm+\left({1616\over
 27}-56\zeta_3\right)\m\nonumber\\
 &&\left.-{1535\over
 12}+{1128\over 9}\zeta_2+{604\over
 9}\zeta_3-{92\over
 5}\zeta_2^2\right\}.
 \end{eqnarray}

 \begin{eqnarray}
 G^{(2),{\rm
 s+v}}_{N,g}&=&C_A^2\left\{\left[32\mm+{44\over
 3}\m\right]\ln^2\left(\frac{Q^2}{\mu^2}\right)\right.\nonumber\\
 &&+\left[-64\mmm-{176\over
 6}\mm-\left({536\over
 9}+112\zeta_2\right)\m-24-{176\over
 3}\zeta_2+24\zeta_3\right]\ln \left(\frac{Q^2}{\mu^2}\right)\nonumber\\
 &&+32\mmmm+{176\over
 9}\mmm+\left({536\over
 9}+112\zeta_2\right)\mm\nonumber\\
 &&+\left({1616\over
 27}-56\zeta_3\right)\m+93+{1072\over
 9}\zeta_2-{308\over
 9}\zeta_3+92\zeta_2^2\nonumber\\
 &&+C_AN_F\left\{\left[-{8\over 3}\m\right]\ln^2
 \left(\frac{Q^2}{\mu^2}\right)+\left[{16\over 3}\mm+{80\over
 9}\m+8+{32\over
 3}\zeta_2\right]\ln \left(\frac{Q^2}{\mu^2}\right)\right.\nonumber\\
 &&\left.-{32\over
 9}\mmm-{80\over 9}\mm-{224\over 27}\m-{80\over
 3}-{160\over
 9}\zeta_2-{88\over 9}\zeta_3\right\}\nonumber\\
 &&+C_FN_F\left\{4\ln \left(\frac{Q^2}{\mu^2}\right)-{67\over
 3}+16\zeta_3\right\}.
 \end{eqnarray}

\end{document}